\documentclass[useAMS,usenatbib,onecolumn]{mn2e}

\usepackage{bm}
\usepackage{calrsfs}
\usepackage{graphicx}

\title[Dynamical friction for accelerated motion]{Dynamical friction for accelerated motion in a gaseous medium}
\author[Fathi Namouni]{Fathi Namouni\thanks{E-mail: namouni@obs-nice.fr}\\
Universit\'e de Nice, CNRS,  Observatoire de la C\^ote d'Azur, BP 4229, 06304 Nice}

\begin{document}

\date{Dates: UNDER REVIEW}

\pagerange{\pageref{firstpage}--\pageref{lastpage}} \pubyear{2009}

\maketitle

\label{firstpage}

\begin{abstract}
Dynamical friction arises from the interaction of a perturber and the gravitational wake it excites in the ambient medium. This interaction is usually derived assuming that the perturber has a constant velocity. In realistic situations, motion is accelerated as for instance by dynamical friction itself. Here, we study the effect of acceleration on the dynamical friction force. We characterize the density enhancement associated with a constantly accelerating perturber with rectilinear motion in an infinite homogeneous gaseous medium and show that dynamical friction is not a local force and that its amplitude may depend on the perturber's initial velocity. The force on an accelerating perturber is maximal between Mach 1 and Mach 2, where it is smaller than the corresponding uniform motion friction. In the limit where the perturber's size is much smaller than the distance needed to change the Mach number by unity through acceleration, a subsonic perturber feels a force similar to uniform motion friction only if its past history does not include supersonic episodes.  Once an accelerating perturber reaches large supersonic speeds, accelerated motion friction is marginally stronger than uniform motion friction. The force on a decelerating supersonic perturber is weaker than uniform motion friction as the velocity decreases to a few times the sound speed. Dynamical friction on a decelerating subsonic perturber with an initial Mach number larger than 2 is much larger than uniform motion friction and tends to a finite value as the velocity vanishes in contrast to uniform motion friction.

\end{abstract}


\section{Introduction}
Dynamical friction is the reaction force that a perturber  feels from the gravitational wake it generates in the ambient medium. This fundamental interaction has therefore had a wide range of applications in the formation of astronomical structures. Applications to collisionless systems include mass segregation in star clusters (e.g. Kim and Morris 2003, Portegies et al 2004), galaxy distribution in hierarchical cluster formation (e.g. Frenk et al 1996, Bullock and Johnston 2005), the decay of satellites in galactic halos and galaxy mergers (Binney and Tremaine 2008 and references therein).  Dynamical friction also occurs in collisional particle systems and was applied to the study of the growth of planetesimals (Hornung et al 1985, Stewart and Whetherill 1988), the eccentricity excitation of planetary embryo orbits (Ida 1990, Kokubo and Ida 1996, Namouni et al 1996), and the confinement of planetary rings (Goldreich and Tremaine 1982). For gaseous systems, applications include the orbital decay of compact stars orbiting supermassive black holes (Narayan 2000), and the heating of intercluster gas by decelerating supersonic galaxies (El-Zant et al. 2004, Faltenbacher et al 2005, Kim et al 2005).  

Following the pioneering work of Chandrasekhar (1943) on the gravitational drag experienced by a perturber in collisionless star systems, the derivations of dynamical friction assume that the perturber has a constant velocity in the ambient medium. The assumption of uniform motion is made for collisionless systems (Binney and Tremaine 2008), collisional particle systems (Hornung et al 1985, Stewart and Whetherill 1988, Ida 1990, Goldreich and Tremaine 1982) as well as  gaseous media (Kim and Kim 2007, Kim et al 2008, Ostriker 1999). The resulting dynamical friction force which depends on the perturber's velocity is subsequently used to model the evolution of the perturber's motion whose velocity varies with time. Only the numerical simulations of the nonlinear fluid equations are free from this assumption (S\'anchez-Salcedo and Brandenburg 1999, 2001).  The neglect of the perturber's acceleration in dynamical friction is justified by the small amplitude of the force that is proportional to the square of the pertuber's mass. Neglecting acceleration is independent of the type of motion and is mainly motivated by the lack of an analytical framework where the perturber's accelerated motion could be taken into account.  To overcome this obstacle, we derive in Appendix A a new expression of dynamical friction that applies to a general type of motion and depends on the perturber's motion and the boundary conditions but not the density perturbation. In particular, dynamical friction is shown to be non-local as it depends on the whole trajectory of the perturber. Although non-locality does not appear if the perturber's velocity is constant, its origin is easy to identify: as the perturber's wake propagates in the medium, a velocity change because of acceleration affects the density enhancement near the perturber and offsets the force from its local value as the far parts of the wake were generated by a different velocity. Therefore when acceleration is taken into account, the amplitude of dynamical friction may change with respect to the value derived assuming a constant velocity because in modelling the force on an accelerated perturber as that on a constant velocity perturber,  one assumes that the gravitational wake produced by accelerated motion is similar to that of uniform motion. We shall see that this difference between the two wakes makes dynamical friction dependent on the initial velocity of the perturber. Another difference between the wakes generated by constant velocity motion and accelerated motion is the absence of a perturbed region with front-back symmetry with respect to a subsonic accelerating perturber. In effect a constant velocity perturber  raises a gravitational wake in the ambient medium that is symmetric with respect to it and truncated by the sonic shockwave. This truncation defines two regions, one that lacks front-back symmetry located behind the perturber,  and one with front-back symmetry around the perturber that becomes the largest region inside the wavefront for velocities smaller  than half the sound speed. Dynamical friction arises from the former region as shown by Ostriker (1999). When acceleration is taken into account, front-back symmetry is lost everywhere. This would intuitively imply that dynamical friction on an accelerated perturber should be much larger than the value derived assuming uniform motion for subsonic perturbers. Curiously, we shall show that the force on an accelerating subsonic perturber initially at rest is similar to the uniform friction force. Lastly, accelerated motion friction would differ from uniform motion friction because the former never reaches steady state. In effect, dynamical friction on a constant subsonic velocity perturber is known to be time-independent (Ostriker 1999).  

It is the object of this paper to quantify how taking into account acceleration modifies dynamical friction on a perturber travelling in an infinite homogeneous gaseous medium. The perturber's motion is assumed to have a constant acceleration, a choice that makes the analytical developments simpler. Physically such constant acceleration approximates the deceleration felt by the perturber  because of dynamical friction as the force amplitude is small and the corresponding instantaneous velocity change is equally small. Constant acceleration also  models the motion of a young star with an asymmetric jet system moving in interstellar gas (Hirth et al 1994, Hartigan and Morse 2007, Namouni 2005,  2007). In what follows, we  characterize the density perturbation associated with accelerated motion in Section 2. We derive the force on an accelerated perturber in  Section 3 using the general expression of dynamical friction (derived in Appendix A) that does not depend on the density enhancement caused by the perturber but only on the perturber's motion and the medium's boundary conditions. In particular we show that including acceleration makes dynamical friction dependent on the initial velocity of the perturber. Section 4 contains concluding remarks. In Appendix B,  the new expression of dynamical friction derived in Appendix A is validated by applying it to uniform rectilinear motion and recovering the known results on dynamical friction.

\section{Density enhancement}
Consider a point-like perturber of mass $M$ on an accelerated rectilinear trajectory, of initial velocity $V_0$ and acceleration $A$,  that travels inside an inviscid isothermal  gaseous medium of density $\rho$ and sound speed $c$. 
Assuming the density perturbation associated with the perturber's wake is small compared to the medium's unperturbed density $\rho_0$, the corresponding density enhancement $\varrho$ defined as $\rho=\rho_0(1+\varrho)$ satisfies the standard forced sound wave equation (Appendix A, equation \ref{wav2A}):
\begin{equation}
\nabla^2\varrho-\frac{1}{c^2}\,
\partial_t^2\varrho=-\frac{{\cal H}(t)}{c^2}\nabla^2\phi_p=-4\pi \frac{GM}{c^2}\, \delta^3[{\bm x}-{\bm \xi}(t)]
{\cal H}(t), \label{waveeq}
\end{equation}
where $\phi_p= -GM/|{\bm x}-{\bm \xi(t)}|$ is the perturber's gravitational potential, ${\cal H}(t)$ is the Heaviside function, $G$ is the gravitational constant, ${\bm \xi}(t)=(At^2/2+V_0t){\bm e}_z$ is the perturber's trajectory chosen along the $z$--direction and $\delta(x)$ is Dirac's delta function.  The solution of this equation is  the Li\'enard-Wiechert potential which may be written in integral form using the retarded Green function as (\ref{dens2A}):
\begin{eqnarray}
\varrho({\bm x},t)
&=&\frac{GM}{c^2}\int_{-\infty}^{+\infty} \frac{\delta
\left[u-t+|{\bm x}-{\bm \xi}(u)|/c\right] {\cal H}(u)}
{|{\bm x}-{\bm \xi}(u)|}\ {\rm d}u. \label{density}
\end{eqnarray}
The acceleration $A$ introduces two natural time and length scales  given by:
\begin{equation}
T=\frac{c}{A}, \ \ L=\frac{c^2}{A}. \label{scalings}
\end{equation}
In these units, the perturber's trajectory may be written as $\xi(t)=t^2/2+{\cal M}_0t$ where ${\cal M}_0$ is the initial Mach number. The Mach number is a function of time and reads: ${\cal M}=t+{\cal M}_0$. The scaling time $T$ is the duration to increase the Mach number by unity and the scaling length $L$ is the corresponding travelled distance.  Applied to the density enhancement these scalings yield:
 \begin{equation}
\varrho({\bm  x},t)=\frac{GMA}{c^4}\int_{-\infty}^{\infty} \frac{\delta
\left(u-t+\left[R^2+\left(z-{\cal M}_0u- u^{2}/2\right)^2\right]^{1/2}\right) {\cal H}(u)}
{\left[R^2+\left(z-{\cal M}_0u- u^{2}/2\right)^2\right]^{1/2}}\ {\rm d}u, \label{density2}
\end{equation}
where  $R^2=x^2+y^2$ is the cylindrical radius; unless it is stated otherwise, all length and time variables and constants are normalized to (\ref{scalings}). In order to derive the density profile we use the property of the $\delta$-function $\alpha(u) \delta[\beta(u)]=\sum_i \alpha(u_i) {\delta(u-u_i)}/{|\beta^\prime(u_i)|}$ 
where $u_i$ are the roots of $\beta(u)$, and $\beta^\prime$ is the derivative of $\beta$. The roots of the argument of the $\delta$--function in equation (\ref{density2}) satisfy the quartic equation:
\begin
{equation}
u^4/4+{\cal M}_0u^3-(1+ z-{\cal M}_0^2)\,u^2+2 (t-{\cal M}_0z) u
-t^2+r^2=0, \label{roots}
\end{equation}
where $r=(R^2+z^2)^{1/2}$ is the spherical radius. This equation is the counterpart of the quadratic equation derived by Ostriker (1999) for uniform motion. The former may be reduced to the latter once time and length dimensions are restored. Equation (\ref{roots}), however, cannot be solved analytically because its varying coefficients do not permit simple and unique expressions of the roots. Instead we determine the roots numerically and substitute them into the density expression obtained by applying the $\delta$--function property cited above. The density expression reduces to:
\begin{equation}
\varrho({\bm  x},t)=\frac{GMA}{c^4}\sum_{{\rm roots}\ u_i {\rm\ of\ eq.\ (\ref{roots})} }|t-u_i-({\cal M}_0+u_i)(z-{\cal M}_0 u_i
- u_i^2/2)|^{-1}. 
\end{equation}
Fig. (1) follows the level curves of $\varrho$ for perturbers initially at rest (${\cal M}_0=0$) as they evolve from $t=0$ to Mach 6. The front-back symmetry that characterizes the density enhancement of subsonic constant velocity motion in a uniform medium is absent for both subsonic and supersonic motion. The perturber reaches Mach 1 at $z=0.5$ but remains inside the sonic wavefront of radius $t$ as the latter has been propagating faster than the perturber's velocity. Once the velocity is supersonic, a Mach cone is formed inside the sonic wavefront and a  shallow density depression develops and trails the perturber ($z\leq -1$). Eventually as Mach 2 is reached the perturber catches up with the sonic shockwave and exits the sonic region to form an external  Mach cone. Fig. (2) shows the level curves of $\varrho$ associated with decelerating perturbers with ${\cal M}_0 \neq 0$ as a function time until the perturber's velocity vanishes. The  density enhancement of subsonic perturbers has a structure similar to that of uniform motion (Ostriker 1999) but lacks its front-back symmetry with respect to the perturber. We show in the following section that under certain circumstances, dynamical friction may be not significantly affected by the loss of front-back symmetry even as closed level curves around the perturber contribute to the force unlike those of uniform motion (see Introduction). The deceleration of supersonic perturbers modifies the structure of the density enhancement and reduces the extent of the Mach cone but does not erase it as the velocity vanishes (bottom row, right panel) indicating the presence of a finite friction force.

\section{Friction force}
The reaction force exerted on a point-like perturber in a homogeneous gaseous medium by its gravitational wake  is given by equation (\ref{for3A}) and is written in dimensional variables as:
\begin{equation}
{\bm F}=\frac{{\cal H}(t)(GM)^2\rho_0}{c^2}\int_{\partial V[{\bm y}+{\bm \xi}(t-r/c)], r\leq ct}r\sin\theta{\rm d}r\, {\rm d}\theta\, {\rm d}\varphi  \ \frac{{\bm y}-{\bm \Delta}}{|{\bm y}-{\bm \Delta}|^3},\label{force}
 \end{equation}
where  $r$, $\theta$ and $\varphi$ are the spherical coordinates of the vector ${\bm y}$ defined as the relative position within the medium with respect to the perturber's position at the retarded time $t-r/c$. It  may be written as ${\bm y}={\bm x}-{\bm \xi}(t-r/c)$ where ${\bm x}$ is the position of fluid element in the medium. This definition of ${\bm y}$ influences the force integral through the boundary conditions of the medium denoted  by $ \partial V$. The phase ${\bm \Delta}={\bm {\bm\xi}}(t)-{\bm {\bm\xi}}(t-r/c)$, the upper radial cutoff at the retarded sonic shockwave $r\leq ct$,  and the boundary conditions are all that is necessary to compute the friction force. In particular, the absence of an explicit expression for the density enhancement is irrelevant for the derivation of the force. The derivation of  expression (\ref{force}) assumes that the perturber is assembled at $t=0$ a fact that introduces the size of the retarded sonic shockwave as an upper cutoff. This assumption is not restrictive as it 
allows us to study the time-dependent evolution of dynamical friction as well approximate steady state behaviour (see section 3.2). It  also describes turning on a numerical simulation of dynamical friction (S\'anchez-Salcedo and Brandenburg 1999, 2001). 

For rectilinear motion along the $z$--direction in an infinite medium, the force is equally along ${\bm e}_z$ and its expression can be reduced further by integrating over the longitude $\phi$ and the variable $\tau=\cos\theta$ as:
\begin{equation}
F=\frac{2\pi{\cal H}(t)(GM)^2\rho_0}{c^2}\int_{\partial V[{\bm y}+{\bm \xi}(t-r)], r\leq t}\ \frac{{\rm d}r}
{\Delta^2}\ \left[\frac{ r-\tau\Delta}{\left|1+(\Delta/r)^2-2(\Delta/r)\tau\right|^{1/2}}\right]^{\tau_{\rm max}}_{\tau_{\rm min}},\label{forcex}
\end{equation}
where  the scalings (\ref{scalings}) have been applied; the perturber's trajectory ${\bm \xi}(t)$, the time, $t$, and the position ${\bm x}$ and its spherical components are dimensionless hereafter. The expressions of $\tau_{\rm min}$ and $\tau_{\rm max}$ as a function of the perturber's trajectory and the variable $r$ are determined from the boundary conditions. For uniform rectilinear motion, ${\cal M}$ is constant and  the phase $\Delta/r={\cal M} $. Calling $r_0$ the gravitational potential's small distance cutoff scaled to $L$ (\ref{scalings}), the integral that appears in the expression (\ref{forcex}) may be written as (Appendix B, equation \ref{ostriA}; Ostriker 1999):
\begin{eqnarray}
I_0&=&{\cal M}^{-2}\left(\frac{1}{2}\left[1+{\cal M}-|1-{\cal M}|\right]\left[1+{\rm sign}(1-{\cal M})\right]-\log \left|\frac{1+{\cal M}}{1-{\cal M}}\right|
-\left[1-{\rm sign} (1-{\cal M})\right] \ \log\left|\frac{(1-{\cal M})t}{r_0}\right|\label{ostriker}\right).
\end{eqnarray}
For accelerated motion, the phase $\Delta/r={\cal M}_0+t-r/2$. In order to examine how acceleration modifies the dynamical friction force, we consider the three cases of perturbers initially at rest (${\cal M}_0=0$),  accelerating perturbers (${\cal M}_0>0$) and decelerating perturbers (${\cal M}_0<0$). 

Before deriving dynamical friction for accelerated motion, a choice of the small distance cutoff $r_0$ must be made. This cutoff is usually taken as the perturber's accretion radius or the system's typical size. The former dimensional cutoff reads  $2GM/V^2$ and is a function of the Mach number whereas the latter is a constant that does not depend on the flow's variables. In this work,  the perturber is treated as a compact object whose gravitational potential is truncated at small distances by the perturber's size. To satisfy such a condition and be consistent with a point-like potential, the perturber's size needs to be  much smaller than the accretion radius $2GM/V^2$. This choice of cutoff  is in agreement with the numerical simulation of the fluid equations in the context of Bondi-Hoyle-Lyttleton accretion (Edgar 2005).  For larger size or extended objects, equation (\ref{force}) no longer applies and dynamical friction may be obtained from the general expression derived for an arbitrary perturbing potential (Appendix A, equation \ref{forc2A}).

\subsection{Perturbers initially at rest}
In an infinite medium, the only relevant physical  boundary condition is that of the small distance cutoff $|{\bm x}-{\bm \xi} (t)|\leq r_0$.\footnote{The variables ${\bm x}$, ${\bm \xi}$ and $r_0$ remain scaled to $L$ (\ref{scalings}).} For a perturber initially at rest, $\xi(t)=t^2/2$, $\Delta/r=t-r/2$ and the Mach number is ${\cal M}=t$. The variables $r$ and $\tau$ therefore need to satisfy the three conditions: (i) $-1\leq\tau\leq 1$, (ii) $\tau\leq [1+(2t-r)^2/4-(r_0/r)^2]/(2t-r)\equiv f(r,t,r_0)$,  and (iii)  $r\leq t$. The last condition implies that $\Delta>0$.  Examination of the three conditions shows $\tau_{\rm min}=-1$,  and  that  $\tau_{\rm max}$ and the remaining bounds on $r$  in the integral (\ref{forcex}) depend on the value of $t$ as the velocity increases. There are five time intervals that need to be considered:
\begin{eqnarray}
\mbox{I}&:& (1+2r_0)^{1/2}-1\leq t\leq 1-(1-2r_0)^{1/2}, \nonumber\\
\mbox{II}&:& 1-(1-2r_0)^{1/2} \leq t \leq 1+(2r_0)^{1/2}, \nonumber\\
\mbox{III}&:& 1+(2r_0)^{1/2} \leq t \leq 1+(1-2r_0)^{1/2}, \nonumber\\
\mbox{IV}&:& 1+(1-2r_0)^{1/2}\leq t\leq 1+(1+2r_0)^{1/2}, \nonumber\\
\mbox{V}&:& 1+(1+2r_0)^{1/2} \leq t. \nonumber
\end{eqnarray}
Case I is that of transitory motion where the density perturbation is in the perturber's vicinity. Prior to $t=(1+2r_0)^{1/2}-1$, the force is zero as the three conditions above cannot be satisfied. The integration domain of $r$ and $\tau_{\rm max}$ are given as:
\begin{eqnarray}
r^f_2\leq r\leq t&\mbox{with}& \tau_{\rm max}=f(r,t,r_0),\ \ \ \mbox{where} \ \ r_2^{f}=1+t-[(1+t)^2-2r_0]^{1/2}. 
\end{eqnarray}
Calling $I$ the integral  in equation (\ref{forcex}) and using these boundaries for $r$, we find:
\begin{eqnarray}
I_{\rm I}&=&\left[ \frac{t^2 r^3-2t^3r^2+(4t^2-4r_0 t+ 2 r_0^2)r - 2 r_0^2 t }{2r_0 t^2r(2t-r)}
-\frac{2t-r_0}{2t^3}\, \log\left|\frac{r}{r-2t}\right| \right] _{r_2^{f}}^{t}, \label{i0}\\
I_{\rm I}&=&-\frac{r_0(2+r^f_2)-(2+r_0+r^f_2)t -r^f_2t^2+t^3}{2r_0t^2}-\frac{2t-r_0}{2t^3}\, \log\left|\frac{2t-r_0}{tr^f_2-r_0}\right|.\label{case1}
\end{eqnarray}
Examination of the variation of $I_{\rm I}$ as a function of time shows that it decays smoothly to zero as $t=(1+2r_0)^{1/2}-1$ is approached. The expression of the force in this time interval is the equivalent of that derived for uniform motion in Appendix B (equation \ref{transient}) when $r_0/|1+{\cal M}|\leq ct\leq r_0/|1-{\cal M}|$  and the wake has not left the perturber's vicinity. In that case, the force read  $F\propto {\cal M}^{-2}\{ -\log[ct(1+{\cal M})/r_0]+{\cal M}-{r_0}/{2ct}+(1-{\cal M}^2)ct/2r_0\}$; it helped explain the transient phase observed before saturation in the numerical simulation of the nonlinear fluid equations (S\'anchez-Salcedo and Brandenburg 1999). 

Case II covers subsonic motion and the corresponding integration boundaries are:
\begin{eqnarray}
r^f_2\leq r\leq r^c_1&\mbox{with}& \tau_{\rm max}=f(r,t,r_0), \ \ \mbox{where}\ \ r_1^{c}=t-1+[(t-1)^2+2r_0]^{1/2}, \label{dom21}\\
r^c_1\leq r\leq  t&\mbox{with}& \tau_{\rm max}=1.\label{dom22}
\end{eqnarray}
The latter domain (\ref{dom22}) does not contribute to the force. The corresponding integrand is proportional to ${\rm sign} (1-t+r/2)-1$  and  the sign argument is positive for $r>2(t-1)$. The expression of $I$ for domain   (\ref{dom21}) may be obtained from equation (\ref{i0}) by substituting $r^c_1$ for $t$. The result is :
\begin{eqnarray}
I_{\rm II}&=&\frac{2r_0 t + (t-r_0) ( r_1^c+r_2^f) -t^2 (r_1^c-r_2^f)}{2 r_0 t^2}- \frac{2t-r_0}{2t^3}\,\log\left|\frac{r_1^ct+r_0}{r_2^ft-r_0}\right|. \label{case2}
\end{eqnarray}
In the limit $r_0\ll 1$ and $t< 1$, $I_{\rm II}(r_0\ll 1)=t^{-2}(2t -\log|(1+t)/(1-t)|)$. This is exactly the  friction force acting on subsonic perturbers with uniform motion given by equation (\ref{ostriker}). It is interesting that although the density enhancements of subsonic accelerated  motion and subsonic uniform motion differ significantly, the corresponding friction forces are identical when the limit $r_0 \ll 1$ is taken for the accelerated motion friction. \footnote{We remind the reader that the uniform motion friction force on subsonic perturbers (\ref{ostriker}) is independent of $r_0$ and that no assumptions on $r_0$ are made to derive its expression.} This result is the more surprising as the density enhancement associated with accelerated motion does not involve a region with front-back symmetry as explained in the Introduction leading one to guess wrongly that friction would be stronger for accelerated motion than it is for uniform motion. The explanation of this agreement comes from the friction force's dependence on the phase $\Delta=\xi(t)-\xi(t-r/c)$ (dimensional variables). As the contribution of the force for subsonic motion comes mostly from the vicinity of the perturber's retarded position ($|{\bm y}|\rightarrow 0$), we have $r/c\ll t$ and the phase  becomes $\Delta=\dot\xi(t)r/c$. The latter is exactly the phase of uniform motion with a Mach number ${\cal M}=\dot \xi(t)/c$ regardless of the type of motion $\xi(t)$ as long as, (1) $r_0\ll 1$ which implies in dimensional variables that $|\ddot\xi(t)|\ll c^2/r_0$, (2) the velocity remains subsonic and (3) the motion does not have a history with supersonic episodes (see section 3.3). 

Case III concerns motion between Mach 1 and Mach 2 where the perturber travels at  supersonic speed but  has not yet crossed the sonic shock wave. The integration domains are given as:
\begin{eqnarray}
r^f_2\leq r\leq r^c_2&\mbox{with}& \tau_{\rm max}=f(r,t,r_0),\ \ \mbox{where} \ \ r_2^{c}=t-1-[(t-1)^2-2r_0]^{1/2},
\label{dom31}\\
r^c_2\leq r\leq  r_3^c&\mbox{with}&  \tau_{\rm max}=1,\ \ \mbox{where}\ \ \ r_3^{c}=t-1+[(t-1)^2-2r_0]^{1/2},\label{dom32}\\
r^c_3\leq r\leq r^c_1&\mbox{with}&  \tau_{\rm max}=f(r,t,r_0),\label{dom33}\\
r^c_1\leq r\leq  t&\mbox{with}& \tau_{\rm max}=1.\label{dom34}
\end{eqnarray}
The integration radii satisfy $r^f_2< r_2^{c}\leq  r_3^{c}\leq 2(t-1)\leq r_1^{c}$. This shows that domain (\ref{dom34}) does not contribute a finite term to the force in the same way as domain (\ref{dom22}).  Domains (\ref{dom31}) and (\ref{dom33}) contribute terms similar to that of equation (\ref{i0}) except that for the former $t$ is replaced with $r^c_2$ and for the latter  $t$ is replaced with $r^c_1$, and $r^f_2$ with $r^c_3$. The second domain (\ref{dom32}) contributes the term $I_{(\ref{dom32})}$ given as:
\begin{eqnarray}
I_{(\ref{dom32})}&=&\left[\frac{4}{(r-2t)t}-\frac{2}{t^2}\log\left|\frac{r}{r-2t}\right|\right]_{r_2^{c}}^{r_3^{c}}, \label{i2}
\end{eqnarray}
which when combined with the contributions of domains (\ref{dom31}) and (\ref{dom33}) yields:
\begin{eqnarray}
I_{\rm III}&=&\frac{2r_0 t -r_0 (r_1^{c}+r_2^{f}-r_2^{c}+r_3^{c}) +t (r_1^{c}+r_2^{f}+r_2^{c}-r_3^{c})}{2 r_0 t^2}-\frac{2t+r_0}{2t^3}\,\log\left|\frac{r_3^{c}t-r_0}{r_2^{c}t-r_0}\right|\nonumber \\
&& -\frac{r_1^{c}-r_2^{f}+r_2^{c}-r_3^{c}}{2 r_0} - \frac{2t-r_0}{2t^3}\,\log\left|\frac{r_1^{c}t+r_0}{r_2^{f}t-r_0}\right|.\label{case3}
\end{eqnarray}
Case IV defines a small time interval near Mach 2 where the perturber exits the sonic wavefront.  The corresponding integration  domains are:
\begin{eqnarray}
r^f_2\leq r\leq r^c_2&\mbox{with}& \tau_{\rm max}=f(r,t,r_0),\\
r^c_2\leq r\leq  r_3^c&\mbox{with}& \tau_{\rm max}=1,\\
r^c_3\leq r\leq t&\mbox{with}& \tau_{\rm max}=f(r,t,r_0).
\end{eqnarray}
The expression of $I$ is obtained by summing up the terms associated with domains (\ref{dom31}) and  (\ref{dom32}) of Case III, and substituting $t$ for $r_1^c$ in domain (\ref{dom33}). The result is:
\begin{eqnarray}
I_{\rm IV}&=&\frac{ r_0(r_2^c-r_3^c-r_2^f-2)+(2 + r_0 + r_2^c - r_3^c + r_2^f)t+( -r_2^c + r_3^c + r_2^f)t^2-t^3}{2 r_0 t^2}\nonumber \\
&&-\frac{2t+r_0}{2t^3}\,\log\left|\frac{r_3^ct-r_0}{r_2^{c}t-r_0}\right| - \frac{2t-r_0}{2t^3}\,\log\left|\frac{2t-r_0}{r_2^{f}t-r_0}\right|. \label{case4}
\end{eqnarray}
Case V concerns supersonic motion that exited the sonic wavefront ${\cal M}>2$. The integration domains are given as:
\begin{eqnarray}
r^f_2\leq r\leq r^c_2&\mbox{with}&  \tau_{\rm max}=f(r,t,r_0),\\
r^c_2\leq r\leq  t &\mbox{with}&  \tau_{\rm max}=1. \label{dom-v}
\end{eqnarray}
The integral $I$ is obtained as previously by combining the domains of Case IV and making the necessary boundary substitutions. The result is:
\begin{equation}
I_{\rm V}=\frac{ r_0 (r_2^{c}-r_2^{f}-4) +(r_2^{c}+r_2^{f})t- (r_2^{c}-r_2^{f}) t^2}{2 r_0 t^2}-\frac{2t+r_0}{2t^3}\,\log\left|\frac{2t+r_0}{r_2^{c}t-r_0}\right| - \frac{2t-r_0}{2t^3}\,\log\left|\frac{2t-r_0}{r_2^{f}t-r_0}\right|. \label{case5}
\end{equation}
For $r_0\ll 1$ and $t>2$, the scaled force reads:
\begin{equation}
I_{\rm V}(r_0\ll 1)=-t^{-2}   \left[2(1+\log 2) +\log\left|\frac{t^2(t^2-1)}{r_0^2}\right|\right].\label{case5l}
\end{equation}
This expression differs from that of equation (\ref{ostriker}) through  the first term in the bracket $-2(1+\log 2)/t^2$  implying that the force on accelerated motion of a perturber initially at rest, is larger than that on uniform motion. Fig. (3) shows the friction force (\ref{case1}), (\ref{case2}),(\ref{case3}), (\ref{case4}), (\ref{case5}) as a function of time, or equivalently the Mach number, for the three values of the small distance cutoff $r_0=10^{-2}$, $10^{-4}$, and $10^{-6}$. The force obtained by using the  uniform motion expression, $I_0$,  (\ref{ostriker}) with ${\cal M}=t$ is also shown for the same values of $r_0$. For subsonic motion, the agreement between the two forces is best when $r_0$ is smallest. The force $I_0$ overestimates maximal drag by about 20\%; it underestimates the drag on supersonic motion by the factor  $\sim 2(1+\log 2)/t^2$. Lastly, we remark that the derivation of dynamical friction in the five time intervals shows that there are no force singularities associated with accelerated motion such as that at Mach 1 for uniform motion (\ref{ostriker}).

\subsection{Perturbers with positive acceleration}
To an accelerating perturber of trajectory, $\xi(t)={\cal M}_0 t +t^2/2$ with ${\cal M}_0>0$, corresponds the phase  $\Delta/r={\cal M}-r/2$ where ${\cal M}=t+{\cal M}_0$ is the Mach number. It could therefore seem sufficient to take the force expressions of perturbers initially at rest and substitute ${\cal M}$ for $t$. There is however a key difference between the two cases: the condition from the retarded sonic wavefront, $r\leq t$, refers to time regardless of the type of motion. It should therefore be replaced with $r\leq {\cal M}-{\cal M}_0$ that we refer to as condition (iv). This difference makes the force dependent on the initial velocity in contrast to the expressions derived assuming a constant velocity (\ref{ostriker}). The physical reason for such a dependence was discussed in the Introduction: as the velocity varies, the far part of the wake that was launched in the medium corresponds to an earlier velocity. Combining the force from the far part of the wake to that near the perturber results in a force that has a memory of the earlier velocity. 

Inspection of the conditions (i,ii) of the previous section along with (iv) shows that the two cases ${\cal M}_0>1$ and ${\cal M}_0<1$ may be considered separately.  In the following we do not derive the force expressions for the counterparts of Cases I and IV as their space extensions vanish when $r_0$ tends to zero.
 
For an accelerated perturber with an initial supersonic speed (${\cal M}_0>1$), the boundary conditions (i, ii, iv) away from the transient phase (Case I) are identical to those of Case V (\ref{dom-v}) except that in $r^f_2$ and $r^c_2$, $t$ is replaced with ${\cal M}$, and the outer integration domain is $r^c_2\leq r\leq {\cal M}-{\cal M}_0$. Provided that
${\cal M}>1+[(1-{\cal M}_0)^2+2r_0]^{1/2}$ that we call Case VI,  the friction force is given as:
\begin{eqnarray}
I_{\rm VI}&=&\frac{r_0 {\cal M}_0( 4 + r^c_2 - r^f_2)+[ {\cal M}_0 (r^c_2 +r^f_2)+ r_0 (r^c_2 - r^f_2-4)]{\cal M} +[r^c_2+r^f_2+{\cal M}_0(r^f_2-r^c_2)]{\cal M}^2+(r^f_2 -r^c_2){\cal M}^3}{2 r_0 {\cal M}^2({\cal M}+{\cal M}_0)} \nonumber \\
&&-\frac{2{\cal M}+r_0}{2{\cal M}^3}\,\log\left|\frac{2{\cal M}+r_0}{r_2^{c}{\cal M}-r_0}\right| - \frac{2{\cal M}-r_0}{2{\cal M}^3}\,\log\left|\frac{2{\cal M}-r_0}{r_2^{f}{\cal M}-r_0}\right| -\frac{2}{{\cal M}^2}\log\left|\frac{{\cal M}-{\cal M}_0}{{\cal M}+{\cal M}_0}\right|.\label{case6}
\end{eqnarray}
New terms depending the initial Mach number appear in the expression of the force. As ${\cal M}_0$ is set to zero the force reduces to (\ref{case5}) with $t={\cal M}$. This shows that including acceleration in the perturber's motion makes the medium store the information about the perturber's initial state which is in turn reflected in the reaction force felt by the perturber.   Taking the limit $r_0\ll 1$, the force becomes:
\begin{eqnarray}
I_{\rm VI}(r_0\ll 1)&=&-\frac{2({\cal M}- {\cal M}_0)}{{\cal M}^2({\cal M}+{\cal M}_0)}-\frac{2\log 2}{{\cal M}^2}  -\frac{2}{{\cal M}^2}\log\left|\frac{{\cal M}-{\cal M}_0}{{\cal M}+{\cal M}_0}\right| -\frac{1}{{\cal M}^2}\,\log\left|\frac{{\cal M}^2({\cal M}^2-1)}{r_0^2}\right|.\label{case60}
\end{eqnarray}
Three extra terms appear with respect to the uniform velocity expression (\ref{ostriker}). As ${\cal M}\gg {\cal  M}_0$, the force tends to  that of Case V (\ref{case5l}) of perturbers initially at rest as the density perturbation of a larger Mach number dominates the initial density perturbation.

For an accelerated perturber with an initial subsonic speed, the boundary conditions (i,ii,iv) show that the force expression is given as:
\begin{eqnarray}
1-\left[(1-{\cal M}_0)^2-2r_0\right]^{1/2}\leq {\cal M}\leq1+(2r_0)^{1/2}:&& I=I_{\rm II},\label{case61}\\
1+(2r_0)^{1/2}\leq {\cal M}\leq 1+\left[(1-{\cal M}_0)^2-2r_0\right]^{1/2}:&& I=I_{\rm III},\label{case62}\\
1+\left[(1-{\cal M}_0)^2+2r_0\right]^{1/2}\leq {\cal M} :&& I=I_{\rm VI},\label{case63}
\end{eqnarray}
where in the expressions of $I_{\rm II}$ and $I_{\rm III}$, $t$ is replaced with ${\cal M}$. We recover the general statement we made about the force on subsonic motion in Case II. In particular, the friction force on an initially subsonic perturber depends on the initial velocity as soon as the velocity is supersonic and becomes larger than $2c-V_0$.  In Fig. (4) we show the friction force with $r_0=10^{-4}$ for an initial subsonic velocity with  ${\cal M}_0=0.6$ using equations (\ref{case61}), (\ref{case62}) and (\ref{case63}), and an initial supersonic velocity with ${\cal M}_0=2$ using (\ref{case6}).

We remark that the dependence of dynamical friction on the initial velocity of the perturber is not an artifact of choosing to assemble the perturber at $t=0$. This starting condition can be replaced with $t=-T_0$ and $T_0$ in turn can be made to tend to infinity thus approximating a steady state. Examination of the general force expression shows that  $T_0 \rightarrow \infty$ is equivalent to the condition $t=0$ where the limits ${\cal M} \gg 1$ and ${\cal M} \ll  1$ are taken for accelerating  and decelerating perturbers respectively. For positively accelerating perturbers (Case IV), the force is independent of the initial Mach number as ${\cal M}\gg1$ (\ref{case60}). In the next section, we show  that for decelerating perturbers the dependence on ${\cal M}_0$ remains as the velocity decays to zero.

\subsection{Decelerating perturbers}
For decelerating perturbers (${\cal M}_0<0$), we seek the dynamical friction force from $t=0$ until the velocity vanishes at $t=|{\cal M}_0|$. The choice of a negative Mach number instead of a negative acceleration makes the derivations easier but results in a positive friction force (opposed to the direction of motion). The phase $\Delta/r={\cal M}-r/2$ is now negative and the boundary conditions that determine the integration domains of (\ref{forcex}) are (i), (vi) and (v)  $\tau\geq f(r,{\cal M},r_0)$. The latter condition shows that $\tau_{\rm max}=1$. Three cases need to be considered: $-1\leq {\cal M}_0$, $-1-(1-2r_0)^{1/2}\leq {\cal M}_0 \leq -1$ and ${\cal M}_0 \leq -1-(1-2r_0)^{1/2}$. 
For  subsonic motion ($-1\leq {\cal M}_0$) and away from the transient regime (i.e. $-1+[(1+{\cal M}_0)^2+2r_0]^{1/2}\leq {\cal M}$), the integration domains are given as:
\begin{eqnarray}
  r^c_1\leq r\leq r^f_2, &\mbox{with}&\tau_{\rm min}=f(r,{\cal M},r_0), \label{decesub1}\\
  r^f_2\leq r\leq {\cal M}-{\cal M}_0,  &\mbox{with}& \tau_{\rm min}=-1.\label{decesub2}
  \end{eqnarray}
As for Case II (equations \ref{dom21} and \ref{dom22}), the latter domain (\ref{decesub2})  does not contribute to the force. The corresponding integrand is proportional to $1-{\rm sign} (1+{\cal M}-r/2)$  and  the sign argument is positive for $r<2({\cal M}+1)$. The contribution of  domain (\ref{decesub1}) is identical to that of Case II and  $I=I_{\rm II}$ (\ref{case2}) with $t$ replaced by ${\cal M}$. Dynamical friction on a decelerating subsonic perturber is therefore independent of its initial velocity. This confirms the general statement we made about subsonic motion in an infinite medium in Case II.

For supersonic perturbers with an initial Mach number $-1-(1-2r_0)^{1/2}\leq {\cal M}_0 \leq -1$,  we need to consider three cases depending on the value of the Mach number ${\cal M}$. The first which we call Case VII corresponds to $-1-[(1+{\cal M}_0)^2-2r_0]^{1/2}\leq {\cal M}\leq -1+(2r_0)^{1/2}$ whose boundaries are:
\begin{eqnarray}
r^c_1\leq r\leq r^f_1, &\mbox{with}& \tau_{\rm min}=f(r,{\cal M},r_0),\ \ \mbox{where}\ \ r_1^{f}={\cal M}+1+[({\cal M}+1)^2+2r_0]^{1/2}, \label{decesup1}\\
r^f_1\leq r\leq {\cal M}-{\cal M}_0, &\mbox{with}& \tau_{\rm min}=-1. \label{decesup2}
\end{eqnarray}
Domain (\ref{decesup1}) contributes a term similar to (\ref{i0}) with $r^c_1$ and $r^f_1$ substituted for $t$ and $r^f_2$ respectively.  Domain (\ref{decesup2})  contributes a term similar to (\ref{i2}) with $r^f_1$ and ${\cal M}-{\cal M}_0$ substituted for $r^c_3$ and $r^c_2$ respectively. The force is then given as:
\begin{eqnarray}
I_{\rm VII}&=&\frac{{\cal M}_0 r_0 (r^f_1- r^c_1-4  )
+ [{\cal M}_0 (r^c_1 +  r^f_1) + r_0 (4 - r^c_1 + r^f_1)] {\cal M}+[r^c_1  + r^f_1 + {\cal M}_0 (r^f_1-  r^c_1)] {\cal M}^2+ ( r^f_1-r^c_1 ){\cal M}^3}{2r_0{\cal M}^2({\cal M}+{\cal M}_0)}+\nonumber\\
&& +\frac{2{\cal M}+r_0}{2 {\cal M}^3}\,\log\left|\frac{2{\cal M}+r_0}{r_1^{f}{\cal M}+r_0}\right| +\frac{2{\cal M}-r_0}{2{\cal M}^3}\,\log\left|\frac{2{\cal M}-r_0}{r_1^{c}{\cal M}+r_0}\right| +\frac{2}{{\cal M}^2}\log\left|\frac{{\cal M}-{\cal M}_0}{{\cal M}+{\cal M}_0}\right| \label{case7}.
\end{eqnarray}
The second case which we call Case VIII concerns $-1+(2r_0)^{1/2}\leq {\cal M}\leq -1+[(1+{\cal M}_0)^2-2r_0]^{1/2}$. The corresponding integration domains are: 
\begin{eqnarray}
r^c_1\leq r\leq r^f_2, &\mbox{with}& \tau_{\rm min}=f(r,{\cal M},r_0),\label{dom81}\\
r^f_2\leq r\leq r^f_3, &\mbox{with}& \tau_{\rm min}=-1, \label{dom82}\\
r^f_3\leq r\leq r^f_1,  &\mbox{with}& \tau_{\rm min}=f(r,{\cal M},r_0),\ \  \mbox{where}\ \ r_3^{f}={\cal M}+1+[({\cal M}+1)^2-2r_0]^{1/2},\label{dom83}\\
r^f_1\leq r\leq {\cal M}-{\cal M}_0, &\mbox{with}& \tau_{\rm min}=-1.\label{dom84}
\end{eqnarray}
The integration radii satisfy $r^c_1< r_2^{f}\leq  r_3^{f}\leq 2({\cal M}+1)\leq r_1^{f}$. In contrast to Case III, this ordering shows that the second domain (\ref{dom82}) does not contribute a finite term to the force. Domain (\ref{dom81}) contributes a term equal to $I_{\rm II}$,  and domain (\ref{dom83}) a term similar to (\ref{i2}) where $r^f_3$ and $r^f_1$ are substituted for $t$ and  $r^f_2$ respectively. The last domain (\ref{dom84}) contributes a term similar to that of (\ref{decesup2}) in Case VII. Summing up all terms, we find: 
\begin{eqnarray}
I_{\rm VIII}&=&
\frac{{\cal M}_0 r_0 (r^f_1 - r^f_2 + r^f_3- r^c_1 -4)+
[{\cal M}_0 (r^c_1 + r^f_1 + r^f_2 - r^f_3) + r_0 (4 - r^c_1 + r^f_1 - r^f_2 + r^f_3)]{\cal M}}{2r_0{\cal M}^2({\cal M}+{\cal M}_0)}+\nonumber\\
&&
+\frac{[r^c_1  + r^f_1 + r^f_2 - r^f_3+ {\cal M}_0 (r^f_1 + r^f_2 - r^f_3 -  r^c_1) ]{\cal M}^2  +( r^f_1 + r^f_2 - r^f_3-r^c_1){\cal M}^3}{2r_0{\cal M}^2({\cal M}+{\cal M}_0)}+\nonumber\\
&& +\frac{2{\cal M}+r_0}{2 {\cal M}^3}\,\log\left|\frac{2{\cal M}+r_0}{r_1^{f}{\cal M}+r_0}\right| + \frac{2{\cal M}-r_0}{2{\cal M}^3}\,\log\left|\frac{2{\cal M}-r_0}{r_3^{f}{\cal M}-r_0}\right| +\frac{2}{{\cal M}^2}\log\left|\frac{{\cal M}-{\cal M}_0}{{\cal M}+{\cal M}_0}\right|-\frac{2{\cal M}-r_0}{2{\cal M}^3}\,\log\left|\frac{r_1^c{\cal M}+r_0}{r_2^f{\cal M}-r_0}\right|. \label{case8}
\end{eqnarray}  
The third case concerns $-1+[(1+{\cal M}_0)^2+2r_0]^{1/2}\leq {\cal M}\leq 0$ whose boundaries turn out to be those of subsonic motion ${\cal M}_0>-1$. The corresponding force is therefore given by (\ref{case2}) where $t$ is replaced ${\cal M}$. Fig. (5, left panel) shows the force on a perturber with an initial velocity ${\cal M}_0=-1.7$ using equations (\ref{case7}), (\ref{case8}) and (\ref{case2}) alongside the uniform motion friction force (\ref{ostriker}). In the supersonic regime the uniform motion force overestimates the drag by about 10\% at its maximum. In the subsonic regime the force  on accelerated motion departs significantly from the  uniform velocity force before joining it at $|{\cal M}|\simeq 2+{\cal M}_0= 0.3$ to assume the expression (\ref{case2}). In particular, when the velocity vanishes so does the force.

Lastly, for a supersonic perturber with  an initial velocity ${\cal M}_0 \leq -1-(1-2r_0)^{1/2}$, we find that: 
\begin{eqnarray}
-1-[(1+{\cal M}_0)^2-2r_0]^{1/2}\leq {\cal M}\leq -1+(2r_0)^{1/2}:&& I=I_{\rm VII},\label{case8x} \\
-1+(2r_0)^{1/2}\leq {\cal M}\leq 0: && I=I_{\rm VIII}.\label{case8x2}
\end{eqnarray}
Fig. (5, right panel) shows the force on a perturber with an initial velocity ${\cal M}_0=-4$ using equations (\ref{case8x}) and (\ref{case8x2}) alongside the uniform velocity friction force (\ref{ostriker}).   In the supersonic regime the uniform motion force overestimates the drag by about 20\% at its maximum. In the subsonic regime the force on accelerated motion departs significantly from that on uniform motion and is finite at ${\cal M}=0$. Taking the limit where $r_0$ is small, we find $I_{\rm VII}(r_0\ll 1)=-I_{\rm VI}(r_0\ll 1)$ and:
\begin{eqnarray}
I_{\rm VIII}(r_0\ll 1)&=&
\frac{4(1+{\cal M}_0+{\cal M})}{{\cal M}({\cal M}+{\cal M}_0)} +\frac{1}{ {\cal M}^2}\,\log\left|\frac{({\cal M}-{\cal M}_0)^2(1-{\cal M})^2}{({\cal M}+{\cal M}_0)^2(1+{\cal M})^3}\right| .
\end{eqnarray}  
At ${\cal M}=0$, $I_{\rm VIII}(r_0\ll 1)=1-4/{\cal M}_0^2$ in contrast to the uniform motion force which vanishes with the perturber's velocity. The knowledge of the history of the perturber's motion is therefore required to evaluate dynamical friction. Simply applying uniform motion dynamical  friction force (\ref{ostriker}) may yield incorrect results regarding the perturber's evolution in the ambient medium as evidenced in Fig. (6) where we plot the relative error between the accelerated motion and uniform motion forces for decelerating supersonic perturbers. A similar conclusion was previously reached by S\'anchez-Salcedo and Brandenburg (1999) who numerically tested the accuracy of the uniform motion friction formula (\ref{ostriker}) using the full nonlinear fluid equations. Following the evolution of a decelerating initial perturber with ${\cal M}_0=1.5$ up to ${\cal M}=0.6$, they  found that equation (\ref{ostriker}) underestimated the friction force as ${\cal M}$ became smaller than unity (see their Fig. 5) which corresponds to our Cases VII and VIII. The authors attributed this discrepancy to the `memory of the wake'.

\section{Concluding remarks}
In this paper, we studied the influence of acceleration on dynamical friction in a gaseous medium. Our main conclusions are stated in the Abstract. This work also constitutes an illustration of the use of the general expression of dynamical friction derived in  Appendix A. In particular, applying that formulation to a perturber with accelerated motion allowed us to illustrate how dynamical friction is not a local force and may depend the perturber's initial state. The meaning of the initial state depends on the problem's configuration. For instance, it may correspond to the time the perturber enters the gaseous medium or to the initial time of a numerical simulation of dynamical friction.  We note that non-locality is not peculiar to rectilinear motion and should be true of the dynamical friction associated with rotational  (circular or eccentric) motion.

The main technical result is the set of expressions $I_{\rm I}$ to $I_{\rm VIII}$ that describe dynamical friction as a function of the initial and current velocities  as the perturber is accelerated (positively or negatively) in the ambient medium. Our derivations have assumed a constant acceleration because the calculations are simpler to carry out in this case. Dynamical friction for constantly accelerating perturbers applies readily to the motion of young stars with asymmetric jet systems.  We believe that the general features we found will hold for a general type of acceleration as long as the instant velocity change remains small. Including the deceleration generated by dynamical friction instead of a constant negative acceleration is likely to further reduce the friction force because the acceleration's amplitude increases with time. 

The dependence of dynamical friction on the perturber's size, $r_0$, is most important for supersonic motion. In the force expressions, the dimensionless radius,  $r_0$,  stands in for the ratio $r_0A/c^2$ of dimensional parameters. The reaction force therefore depends explicitly on acceleration even in the limit $r_0\ll c^2/A$ as in  Case VI.  Dynamical friction on subsonic motion is independent of $r_0$ when $r_0\ll c^2/A$. Using the force expression requires the knowledge of the perturber's acceleration and the medium's sound speed. In Fig. (3) to (5), the value $r_0=10^{-4}$ was used so as to make accelerated motion friction closer  to uniform motion friction (see Fig. 1).

The assumption of constant velocity in the derivation of dynamical friction is also made for collisionless star systems. Although they have different magnitudes, the uniform motion friction forces in a gaseous medium and a collisionless star system behave similarly (as $|F|\propto {V}^{-2}$ for supersonic motion and $|F|\propto {\ V}$ for subsonic motion; see  Ostriker 1999, S\'anchez-Salcedo and Brandenburg 2001). This similarity suggests that the conclusions we found are likely to hold for collisionless systems.

\section*{Acknowledgments}

The author thanks an anonymous referee and Scott Tremaine for useful comments.

\appendix \section{General expression of dynamical friction}
In this appendix, we derive a general expression of dynamical friction on a perturber of gravitational potential $\phi_p$  with a general trajectory ${\bm \xi}(t)$  that travels inside an inviscid homogeneous gaseous medium.  The perturber is set in motion at time $t=0$. The medium's response to the perturber's presence is described by the standard fluid equations:
\begin{equation}
\partial_t\rho+{\bm \nabla} \cdot (\rho {\bm v})=0,\ \ \ \partial_t {\bm v}+({\bm v}\cdot{\bm \nabla}) {\bm v}=-\frac{1}{\rho}{\bm \nabla} p-{\cal H}(t){\bm \nabla}{\phi_p},
\end{equation}
where ${\bm v}({\bm x},t)$ and  $\rho({\bm x},t)$ are the perturbed velocity and density fields, $p$ is the pressure, and ${\cal H}(t)$ is the Heaviside function. Prior to the perturber's motion, the gaseous medium is assumed to be at rest $[i.e. {\bm v}({\bm x},t)=0]$, homogeneous of density $\rho_0$, pressure $p_0$ and isothermal with a sound speed $c=(p_0/\rho_0)^{1/2}$. The linearization of medium's equation in the vicinity of the rest state yields: 
\begin{equation}
\partial_t\varrho+c{\bm \nabla} \cdot { \bm{\nu}}=0,\ \ \ 
\frac{1}{c}\partial_t {\bm \nu}+{\bm \nabla} \varrho= -\frac{{\cal H}(t)}{c^2}{\bm \nabla} \phi_p,  \label{momentumeq}
\end{equation}
where we used $\rho=\rho_0(1+\varrho)$ and ${\bm v}=c {\bm \nu}$. 
The combined equations in turn yield the forced sound wave equation satisfied by the density perturbation $\varrho$ as: 
\begin{equation}
\nabla^2\varrho-\frac{1}{c^2}\,
\partial_t^2\varrho=-\frac{{\cal H}(t)}{c^2}\nabla^2\phi_p \label{wavA}.
\end{equation}
The solution of this equation is known in electromagnetism theory as the Li\'enard-Wiechert potential (Jackson 1998, Chapter 14) and represents the electric potential of a moving electric charge distribution.  It is written in integral form using the retarded Green function as:
\begin{eqnarray}
\varrho({\bm x},t)&=&\frac{1}{4\pi c^2}\int_{-\infty}^{+\infty}{\rm d}^3x^\prime {\rm d}u\ \frac{\delta
\left[u-t+|{\bm x}-{\bm x}^\prime|/c\right] \ \nabla^2\phi_p[{\bm x}^\prime - {\bm \xi}(u)]\ {\cal H}(u)}
{|{\bm x}-{\bm x}^\prime|}. \label{densA}
\end{eqnarray} 
In reaction to the perturber's  excitation of the wake, the gaseous medium exerts a force on the perturber given as:
\begin{equation}
{\bm F}=\int_{\partial V({\bm x})} \rho_0\varrho {\cal H}(t) {\bm \nabla} \phi_p {\rm d}^3x =\frac{{\cal H}(t)\rho_0}{4\pi c^2}
\int_{\partial V({\bm x})}{\rm d}^3x{\rm d}^3x^\prime {\rm d}u\  \frac{\delta
\left[u-t+|{\bm x}-{\bm x}^\prime|/c\right]\  \nabla^2\phi_p[{\bm x}^\prime - {\bm \xi}(u)]\ {\bm \nabla} \phi_p[{\bm x} - {\bm \xi}(t)]\ {\cal H}(u)}
{|{\bm x}-{\bm x}^\prime|}\label{forc1A}
\end{equation}
The symbol $\partial V({\bm x})$ denotes the boundaries of the spacial integration domain defined by the medium's physical boundary as well as the lower distance cutoff in the perturber's vicinity such as its size or its corresponding accretion radius. The boundaries  $\partial V({\bm x})$ are used to truncate the perturber's potential at small and large distances to avoid divergence.
Applying the variable changes, ${\bm X}={\bm x}-{\bm \xi}(t)$ and  ${\bm Y}={\bm x}-{\bm x}^\prime$ and integrating the force over $u$ yields the general expression of dynamical friction as:
 \begin{eqnarray}
{\bm F} &=&\frac{{\cal H}(t)G\rho_0}{c^2}
\int_{\partial V[{\bm X}+{\bm \xi}(t)], |{\bm Y}|\leq ct}{\rm d}^3X\frac{{\rm d}^3Y}{|{\bm Y}|}  \  \ \rho_p({\bm X}-{\bm Y}+{\bm \Delta})\ {\bm \nabla} \phi_p({\bm X}),
\label{forc2A} \end{eqnarray}
where $\rho_p$ is the perturber's density distribution, $\rho_p=\nabla^2\phi_p/4\pi G$, and ${\bm \Delta}={\bm \xi}(t)-{\bm \xi}(t-|{\bm Y}|/c)$.  For a given perturbing potential, dynamical friction therefore depends  only on the phase ${\bm \Delta}$ as well as the boundary conditions specified by $\partial V$ and $|{\bm Y}|\leq ct$. The dependence of ${\bm \Delta}$ on the whole trajectory ${\bm \xi}(t)$ shows that dynamical friction is manifestly non-local. Deriving dynamical friction through equation (\ref{forc2A}) is in contrast with the previous two-step methods of estimating dynamical friction for uniform motion by first calculating explicitly  (\ref{densA}) then applying an analytical or numerical calculation to the first integral in equation (\ref{forc1A}) to derive the force (Ostriker 1999, Kim and Kim 2007, Kim et al 2008). As a result such methods obscure the true dependence of the force on the perturber's motion and the boundary conditions. For a point-like perturber, $\phi_p(\bm x)=-GM/|{\bm x}|$ and $\rho_p ({\bm x}) =M \delta({\bm x})$, where $M$ is the perturber's mass. The forced sound wave equation, density enhancement and friction force are reduced to: 
\begin{eqnarray}
&&\nabla^2\varrho-\frac{1}{c^2}\,
\partial_t^2\varrho=-\frac{1}{c^2}\nabla^2\phi=-4\pi \frac{GM}{c^2}\, \delta^3[{\bm x}-{\bm \xi}(t)]
{\cal H}(t), \label{wav2A}\\
\varrho({\bm x},t)
&=&\frac{GM}{c^2}\int_{-\infty}^{+\infty} \frac{\delta
\left[u-t+|{\bm x}-{\bm \xi}(u)|/c\right] {\cal H}(u)}
{|{\bm x}-{\bm \xi}(u)|}\ {\rm d}u, \label{dens2A}\\
{\bm F}&=&\frac{{\cal H}(t)(GM)^2\rho_0}{c^2}\int_{\partial V[{\bm y}+{\bm \xi}(t-r/c)], r\leq ct}r\sin\theta{\rm d}r\, {\rm d}\theta\, {\rm d}\varphi  \ \frac{{\bm y}-{\bm \Delta}}{|{\bm y}-{\bm \Delta}|^3},\label{for3A}
 \end{eqnarray}
where ${\bm y}={\bm x}-{\bm \xi}(t-r/c)$. The condition on the time of perturbation ($t\geq 0$) represented by the Heaviside function in equation (\ref{wavA}) and the expression (\ref{forc1A}) now implies that the spacial domain that enters the expression of the force is equal to the size of the wavefront, $ct$. However, the spacial domain refers to the relative position with respect to the perturber not at time $t$ but at the retarded time $t-r/c$. This explains why the force (\ref{for3A}) describes subsonic as well as supersonic motion that perturbs the medium outside the sonic wavefront of radius $ct$.  In the force expression, the boundaries $\partial V[{\bm y}+{\bm \xi}(t-r/c)]$  are treated as  truncation boundaries. In the next Appendix, the force expression is applied to a point-like perturber in uniform rectilinear motion to recover the corresponding known results.

 \section{Application to uniform rectilinear motion}

We validate the expressions (\ref{forc2A}) and (\ref{for3A}) by deriving dynamical friction for a perturber set in uniform rectilinear motion in an infinite homogeneous medium. The perturber's trajectory is written as ${\bm \xi}(t)=Vt {\bm e}_z$ where ${\bm e}_z$ is the unit vector along $z$, the phase ${\bm \Delta}={\cal M} r {\bm e}_z$ where ${\cal M}$ is the Mach number. The friction force is equally along  ${\bm e}_z$ and its expression is simplified as follows:
\begin{eqnarray}
{F}&=&  \frac{{\cal H}(t)(GM)^2\rho_0}{c^2}\int_{\partial V[{\bm y}+{\bm \xi}(t-r/c)], r\leq ct}r\sin\theta{\rm d}r\, {\rm d}\theta\, {\rm d}\varphi  \ \frac{r\cos \theta-{\Delta}}{|r^2+\Delta^2-2r\Delta\cos\theta|^{3/2}},\\
F&=&\frac{2\pi {\cal H}(t)(GM)^2\rho_0}{V^2}\int_{\partial V[{\bm y}+{\bm \xi}(t-r/c)], r\leq ct} \ \frac{{\rm d}r}
{r} \left[\frac{1-{\cal M}\tau}{({1+{\cal M}^2-2 {\cal M}\tau})^{1/2}}\right]_{\tau_{\rm min}}^{\tau_{\rm max}}, \label{forc3A}
\end{eqnarray}
where in the last expression we performed the integrals over $\varphi$ and $\tau=\cos\theta$.  The extremal values $ \tau_{\rm min}$ and $\tau_{\rm max}$ are determined from the boundary conditions.  Calling $r_0$ the perturber's size, the force is truncated at small distances by $|{\bm x}-{\bm\xi}(t)|>r_0$. Expressing these conditions in terms of the variables $r$ and $\tau$  that appear in equation (\ref{forc3A}) shows that we need to find the domains that satisfy: (i) $-1\leq\tau\leq1$, (ii) $r_0\leq r(1+{\cal M}^2-2 {\cal M}\tau)^{1/2}$, and (iii) $r\leq ct$. The first two conditions yield the following integration domains:
\begin{eqnarray}
\frac{r_0}{|1+{\cal M}|}  <r<\frac{r_0}{|1-{\cal M}|} &\mbox{with}& \tau_{\rm min}=-1,\ \ \ \tau_{\rm max}=\frac{1+{\cal M}^2-(r_0/r)^2}{2{\cal M}}, \label{dom1A}\\ 
\frac{r_0}{|1-{\cal M}|}  <r \ \ \ \ \ \ \ \ \ \ \ \ \ \ \ &\mbox{with}& \tau_{\rm min}=-1,\ \ \ \tau_{\rm max}=1.\label{dom2A}
\end{eqnarray}
The inner domain (\ref{dom1A}) corresponds to the vicinity of the perturber's retarded position. The radius $r_0/|1+{\cal M}|$ defines a region around the perturber that does not contribute to dynamical friction as the first two conditions are not satisfied. The third condition depends on the distance $ct$ travelled by the retarded wavefront.  There are two force regimes: the local perturbation regime  ($ct\leq r_0/|1-{\cal M}|$) where the perturbation remains in the perturber's vicinity and the  distant perturbation regime ($r_0/|1-{\cal M}|\leq ct$ ) where the perturbation has expanded beyond the perturber's vicinity.  Calling $I$ the integral that appears in (\ref{forc3A}),  the integration for the distant perturbation regime extends over the two domains (\ref{dom1A}) and (\ref{dom2A}) with $r \leq ct$ and $I$  is given as:
\begin{eqnarray}
I&=& \int^{\frac{r_0}{|1-{\cal M}|}}_{\frac{r_0}{|1+{\cal M}|}} \left[\frac{1}{2r_0}\left(1-{\cal M}^2+\frac{r_0^2}{r^2}\right)-\frac{1}{r}\right]\ {\rm d}r-\left[1-{\rm sign} (1-{\cal M})\right]\int_{\frac{r_0}{|1-{\cal M}|}}^{ct}\frac{{\rm d}r}{r},\\I&=&-\log \left|\frac{1+{\cal M}}{1-{\cal M}}\right|+\frac{1}{2}\left[1+{\cal M}-|1-{\cal M}|\right]\left[1+{\rm sign}(1-{\cal M})\right]
-\left[1-{\rm sign} (1-{\cal M})\right] \ \log\left|\frac{(1-{\cal M})ct}{r_0}\right|. \label{ostriA}
\end{eqnarray}
This expression agrees with those derived by Ostriker (1999) for subsonic and supersonic perturbers moving in an infinite gaseous medium using explicit expressions of the density perturbation.  In particular,  the force, $F\propto I/{\cal M}^2$,  is time-independent and  behaves as $-{\cal M}$ for small Mach numbers. We note the interesting result that only the vicinity of a subsonic perturber's retarded position (\ref{dom1A}) contributes to dynamical friction. We show in section 3 that this property implies that the force on subsonic perturbers with a general type of motion (not necessarily rectilinear or uniform) is similar to (\ref{ostriA}) provided that the perturber's motion history does not involve supersonic episodes  and that the perturber's acceleration  $|\ddot {\bm \xi}(t)|$ is much smaller than  $c^2/r_0$. 

For the local perturbation regime ($ct\leq r_0/|1-{\cal M}|$),  the force is zero before $t=r_0/c|1+{\cal M}|$ and only the domain (\ref{dom1A}) is relevant to the integral $I$ which is written as:
\begin{eqnarray}
I&=&  \int^{ct}_{\frac{r_0}{|1+{\cal M}|}} \left[\frac{1}{2r_0}\left(1-{\cal M}^2+\frac{r_0^2}{r^2}\right)-\frac{1}{r}\right]\ {\rm d}r= -\log\left[\frac{ct(1+{\cal M})}{r_0}\right]+{\cal M}-\frac{r_0}{2ct}+\frac{(1-{\cal M}^2)ct}{2r_0}. \label{transient}
\end{eqnarray}
In Fig. (7), we show the friction force as a function of $t$ for subsonic and supersonic perturbers as $t$ increases smoothly from 0 to a value in the distant perturbation regime. The time dependence of the force agrees with  that obtained from the numerical simulation of the nonlinear fluid equations (S\'anchez-Salcedo and Brandenburg 1999, Figure 2). However, the attempted fit of the force in the local perturbation regime from numerical simulations  (S\'anchez-Salcedo and Brandenburg 2001, Equation 22) does not reproduce the correct dependence of (\ref{transient}) on ${\cal M}$, $r_0$ and time.

\newpage
\begin{figure}
\begin{center}
\includegraphics[width=65mm]{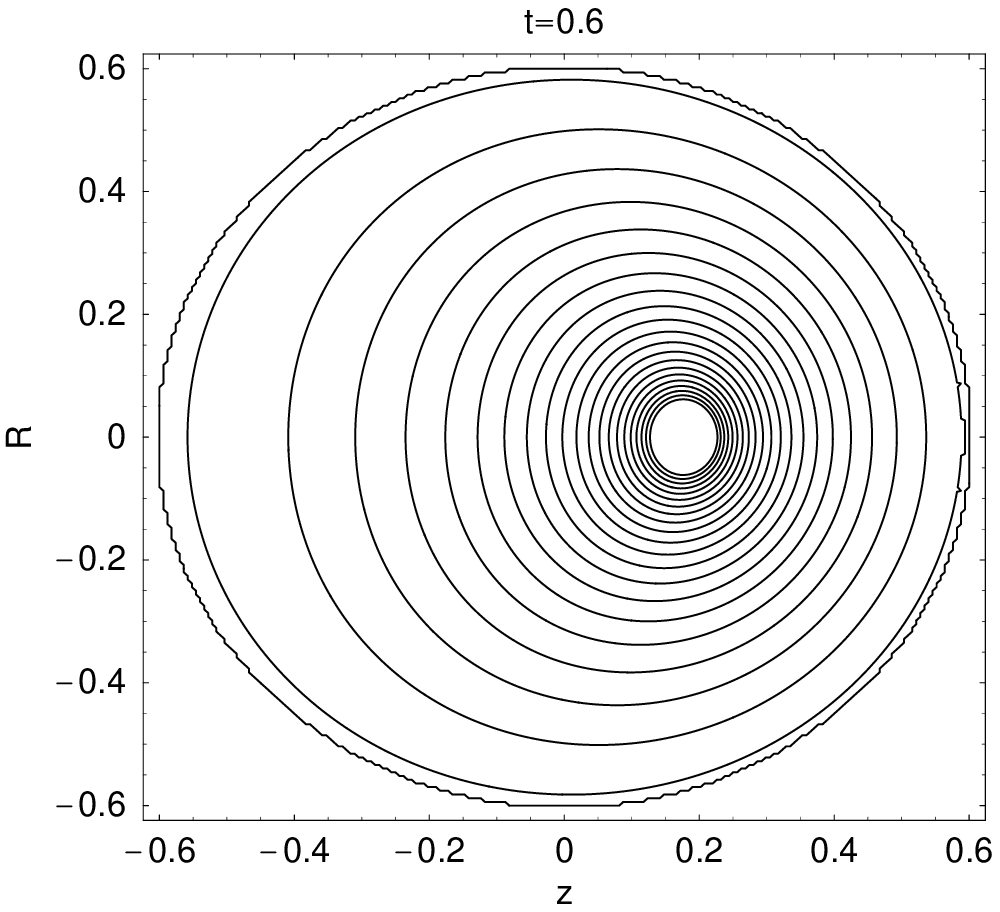}\includegraphics[width=65mm]{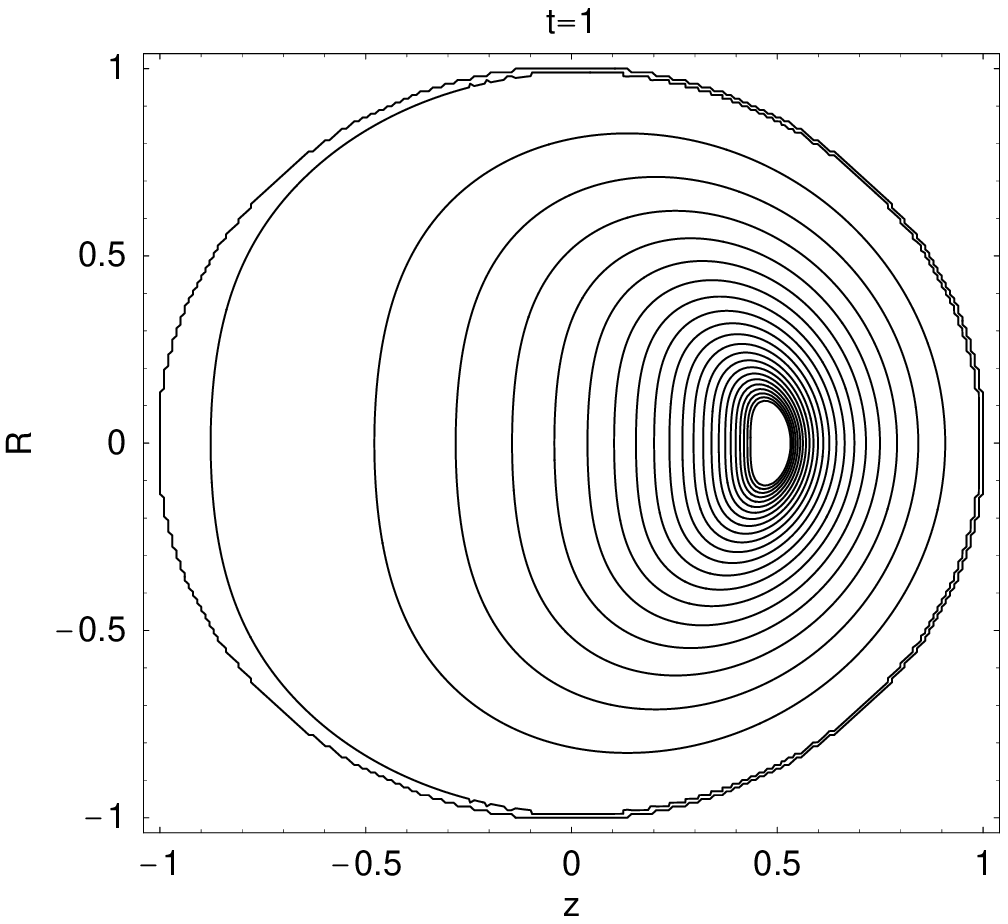}\\
\includegraphics[width=65mm]{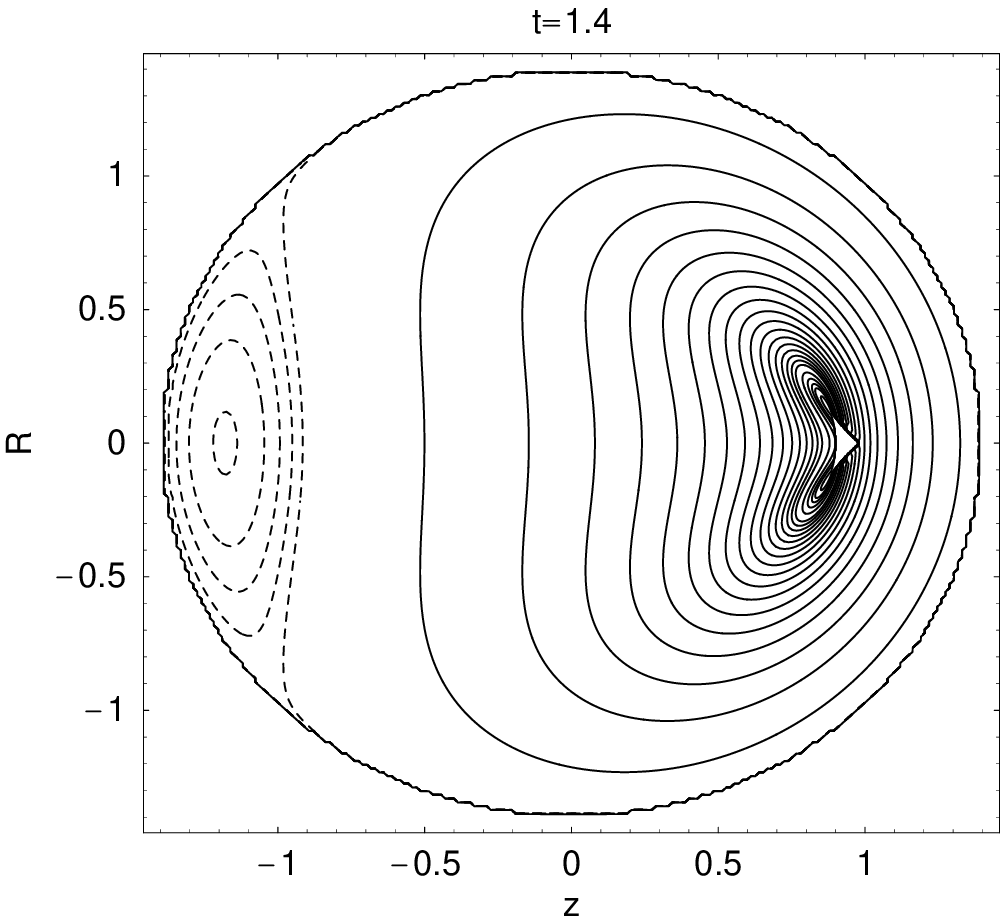}\includegraphics[width=65mm]{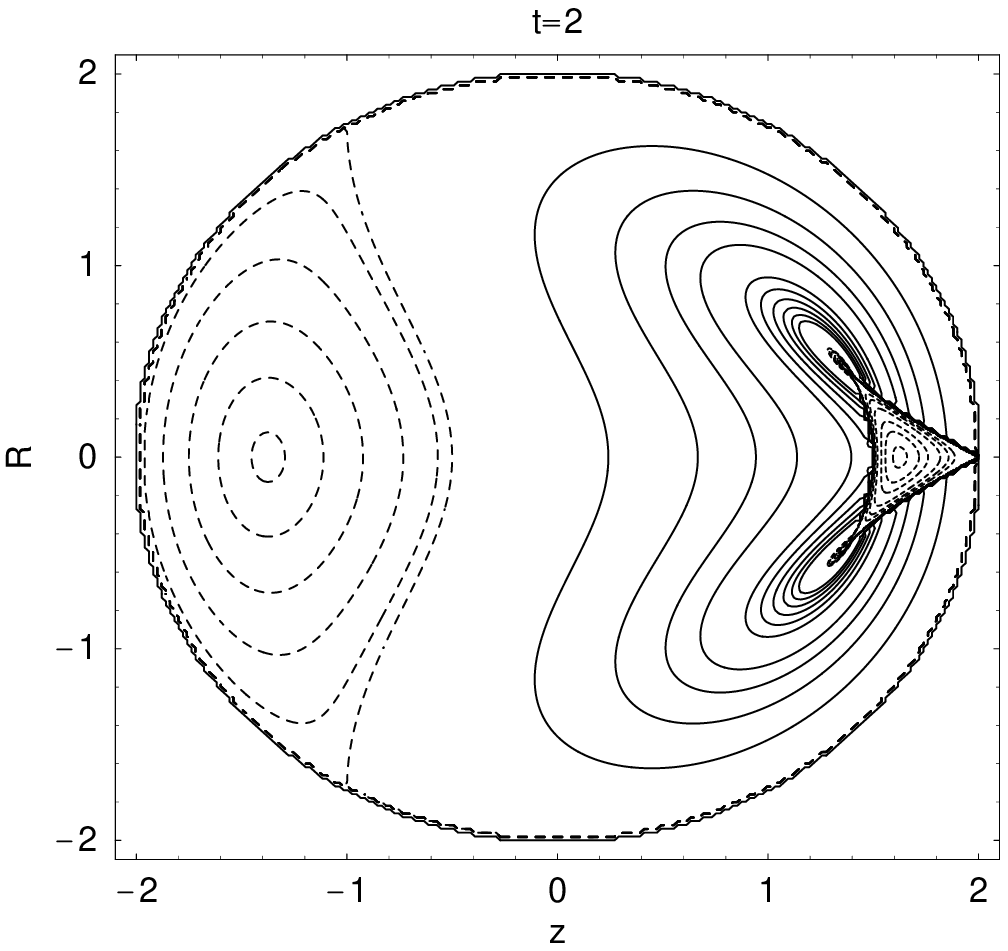}\\
\includegraphics[width=65mm]{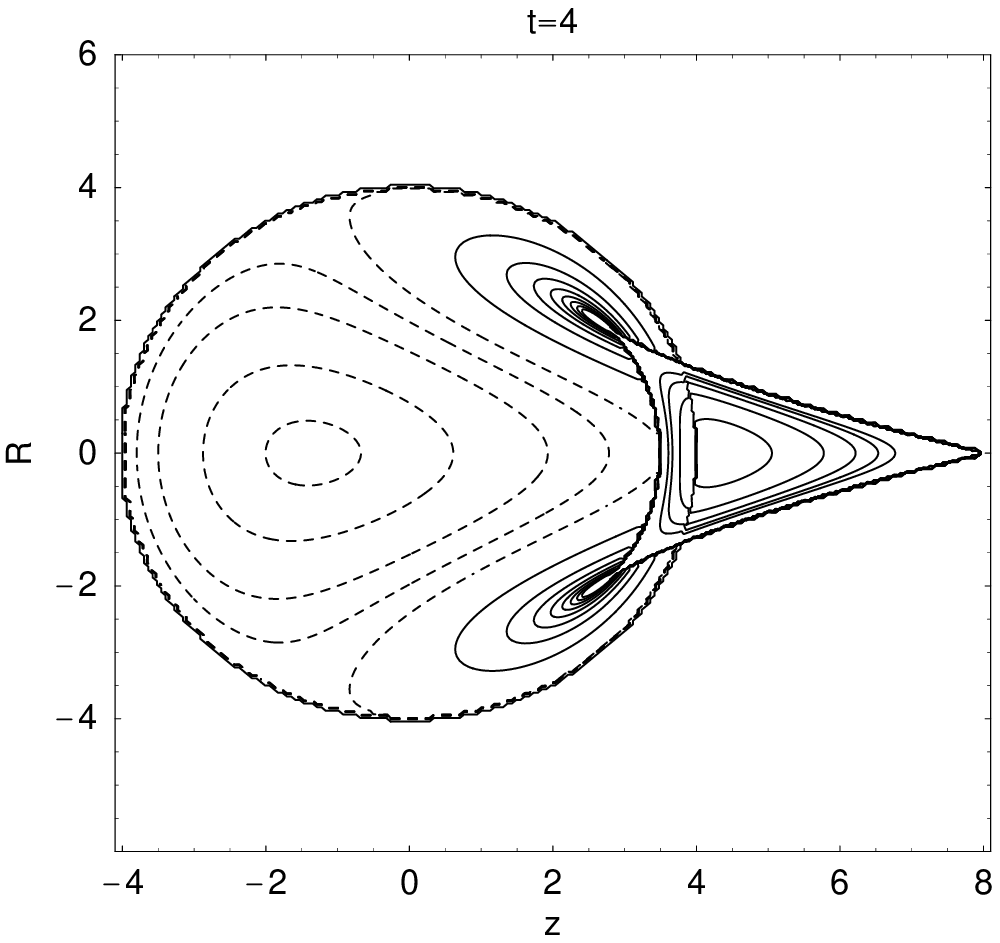} \includegraphics[width=65mm]{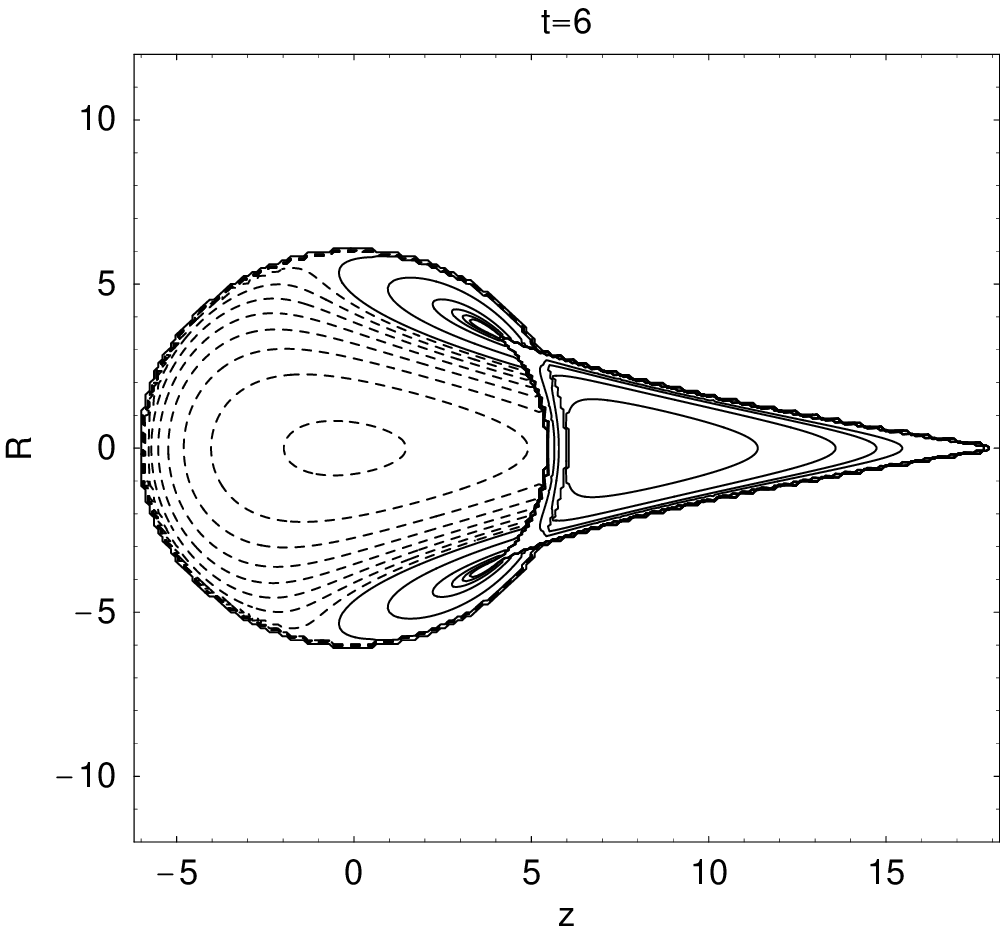}
\end{center}
\caption{Evolution of the density enhancement $\varrho$ from a perturber initially at rest. 
The level curves are determined for values of  $\bar\varrho=\log(1+c^4\varrho/GMA)$. Top left panel: subsonic motion, ($1\leq \bar\varrho\leq 3$, $\Delta\bar\varrho=0.1$). The density increases nearer the perturber. Note how the level curves do not possess front-back symmetry with respect to the perturber. Top right panel: motion at Mach 1. Middle left panel: supersonic motion inside the sonic wavefront. The solid level curves correspond to  ($0.5\leq \bar\varrho\leq 2$, $\Delta\bar\varrho=0.1$) and    ($2<\bar\varrho\leq 3$, $\Delta\bar\varrho=0.2$) near the perturber's Mach cone. The dashed level curves correspond to a shallow depression around $\bar\varrho=0.527$.  Middle right panel: the perturber reaches the sonic shockwave.  The solid level curves correspond to  ($0.5\leq \bar\varrho\leq 2$, $\Delta\bar\varrho=0.1$). The dashed level curves correspond to shallow depressions around $\bar\varrho=0.37$ inside the sonic wavefront and $\bar\varrho=2.3$ inside the Mach cone. Bottom left panel: supersonic motion at ${\cal M}=4$. The solid level curves correspond to  ($0.6\leq \bar\varrho\leq 1$, $\Delta\bar\varrho=0.1$). The dashed level curves correspond to a shallow depression around $\bar\varrho=0.22$.  Bottom right panel: supersonic motion at ${\cal M}=8$. The solid level curves correspond to  ($0.2\leq \bar\varrho\leq 0.6$, $\Delta\bar\varrho=0.1$). The dashed level curves correspond to a shallow depression around $\bar\varrho=0.08$.   For supersonic motion, larger values of $\bar\varrho$ are confined to the sides of the Mach cone where the discontinuity with the unperturbed medium is largest. 
} 
\label{f1}
\end{figure}

\begin{figure}
\begin{center}
\includegraphics[width=65mm]{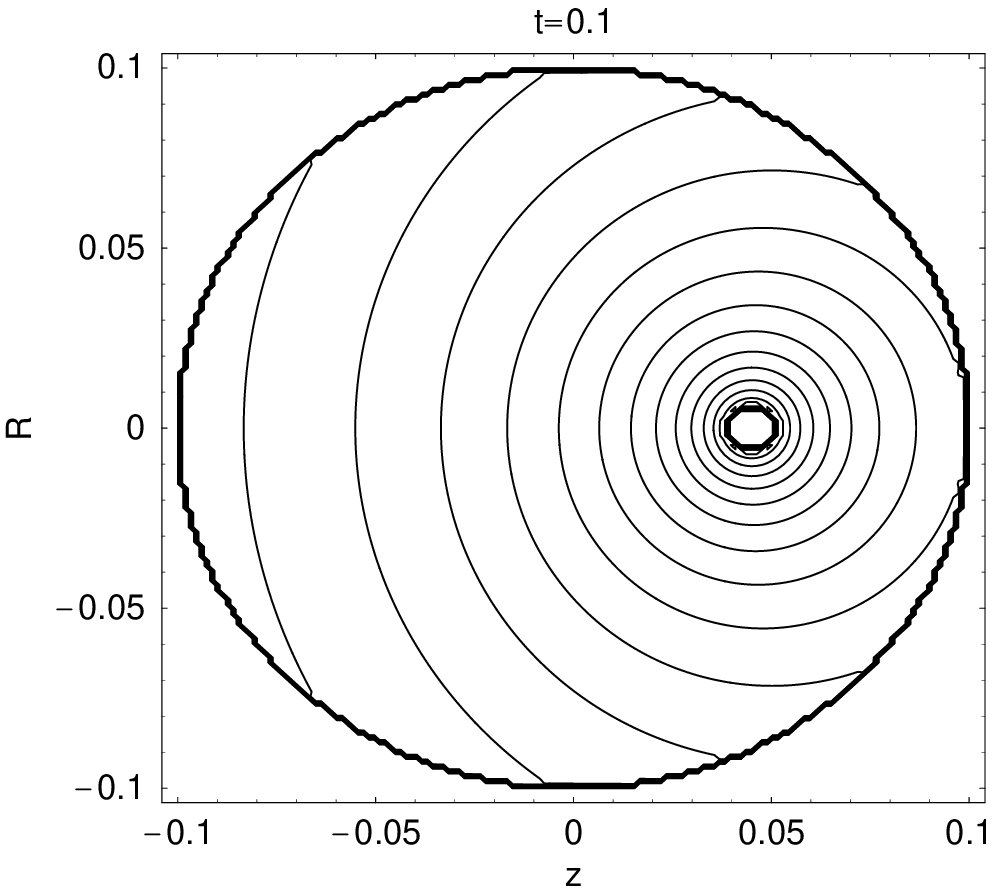}\includegraphics[width=65mm]{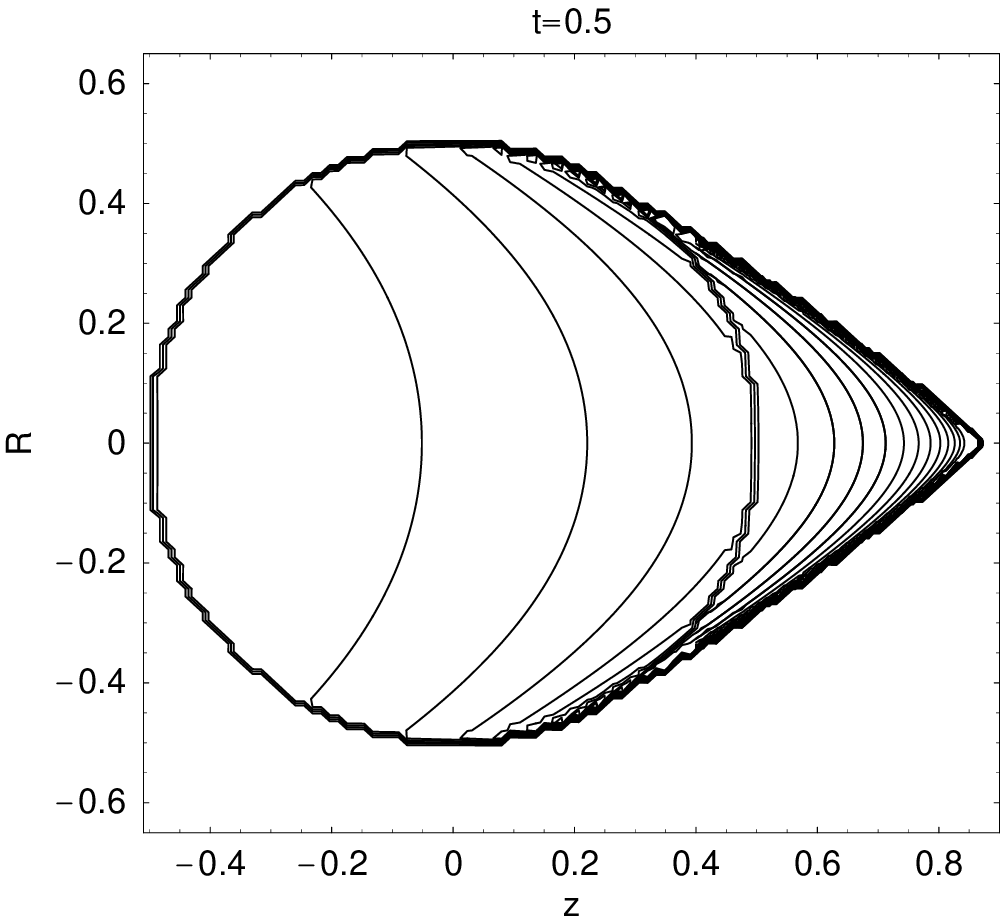}\\
\includegraphics[width=65mm]{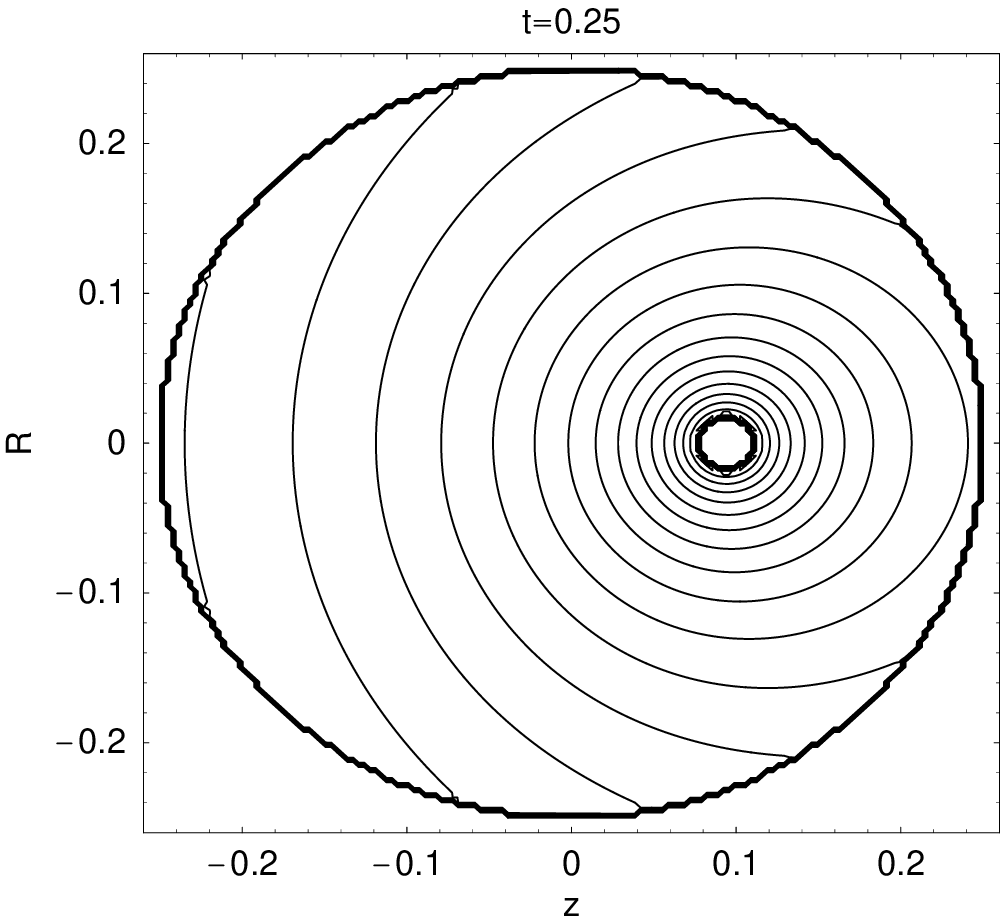}\includegraphics[width=65mm]{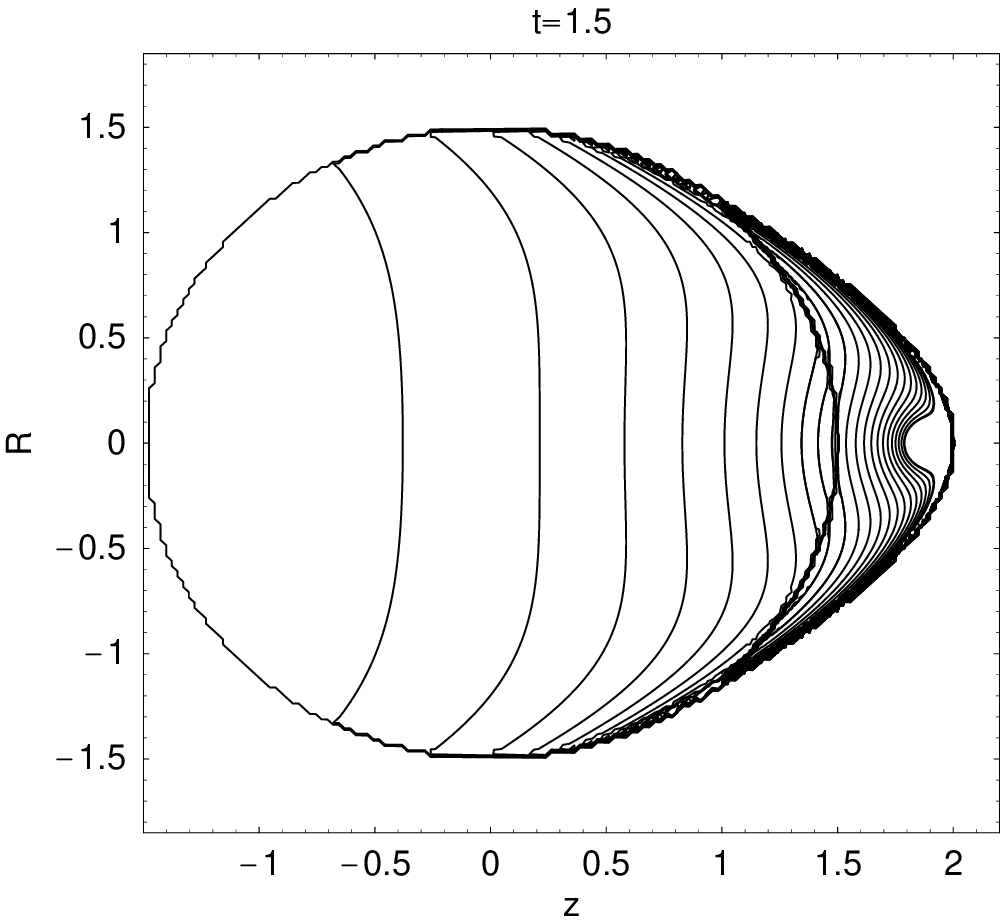}\\
\includegraphics[width=65mm]{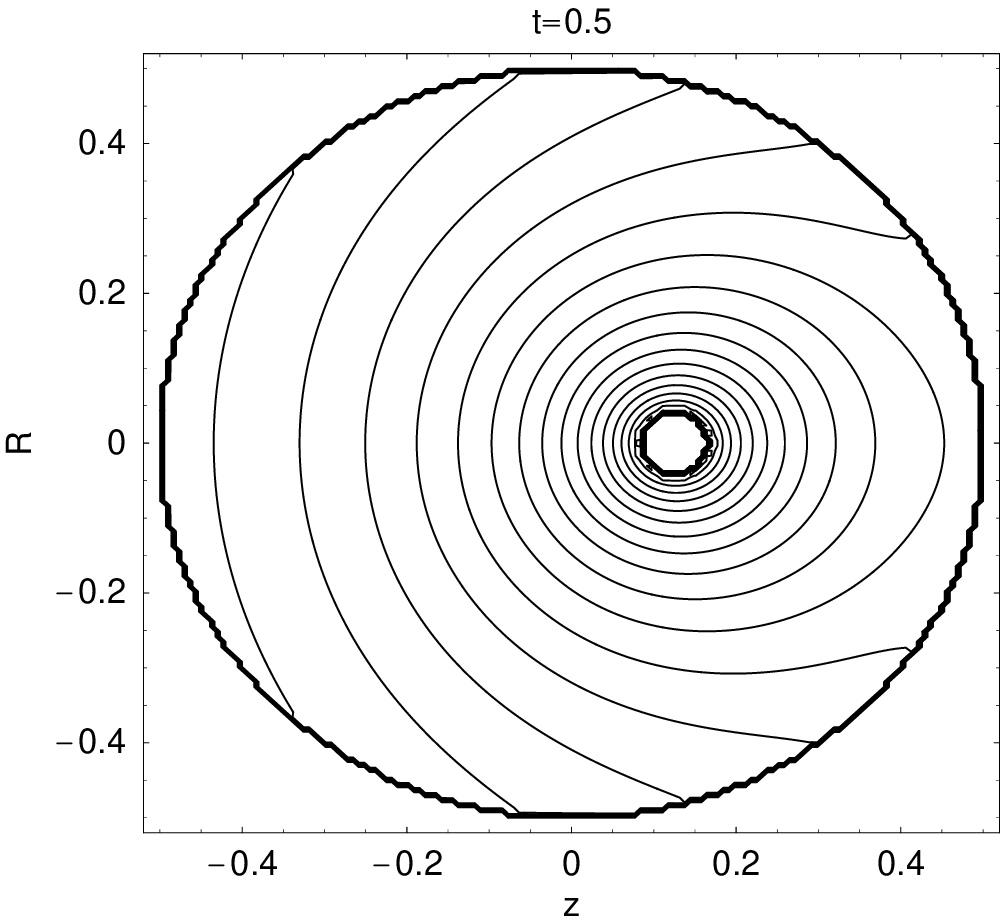} \includegraphics[width=65mm]{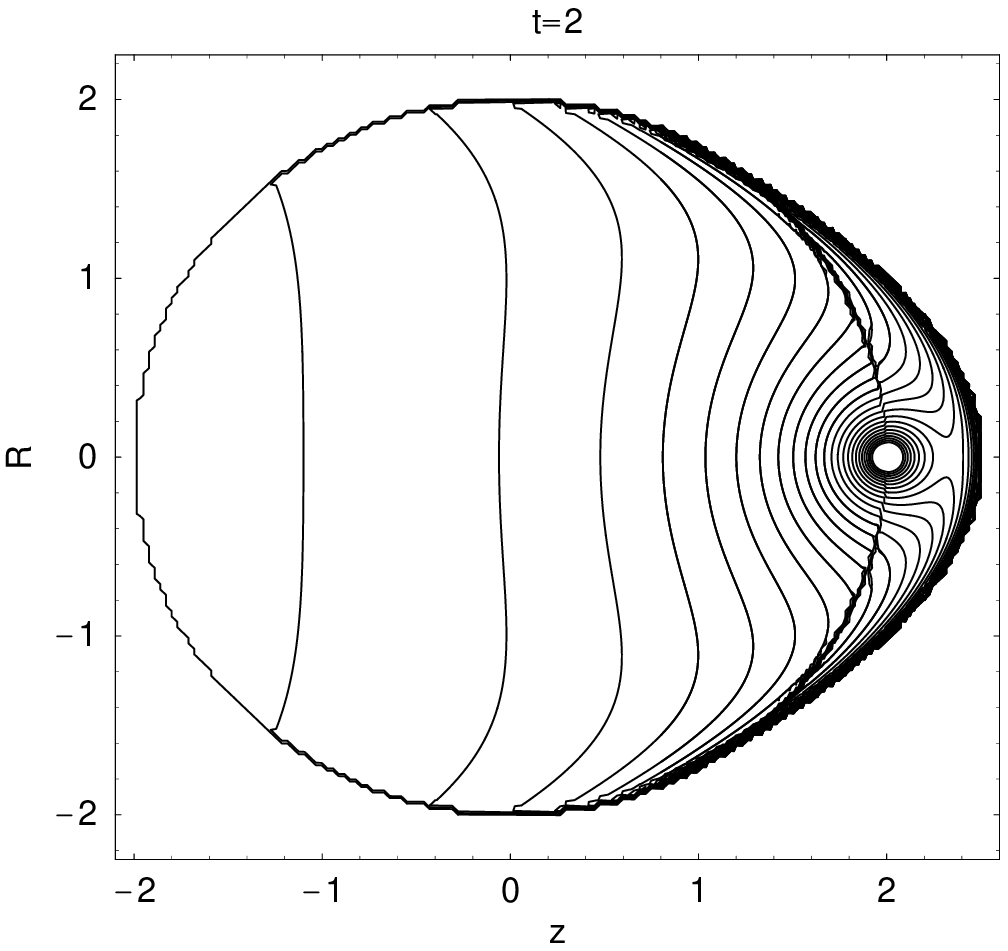}
\end{center}
\caption{Evolution of the density enhancement $\varrho$ from decelerating perturbers. The level curves are shown for a subsonic perturber with 
${\cal M}_0=-0.5$ (left column) and a supersonic perturber with ${\cal M}_0=-2$ (right column). Note how front-back symmetry is lost everywhere for the subsonic perturber. This however does not significantly modify dynamical friction from its value obtained using a constant velocity. }
\label{f2}
\end{figure}

\begin{figure}
\begin{center}
\includegraphics[width=85mm]{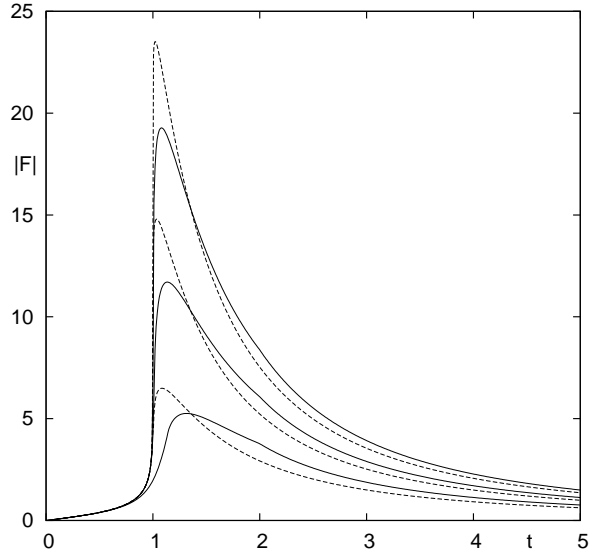}
\end{center}
\caption{Friction force on an accelerating  perturber initially at rest for three values of $r_0=10^{-2}$,  $10^{-4}$, and $10^{-6}$. The solid curves correspond to the friction force that takes acceleration into account given by equations (\ref{case1}), (\ref{case2}), (\ref{case3}), (\ref{case4}) and (\ref{case5}), and the dashed curves to the uniform motion friction force expression (\ref{ostriker}).}
\label{f3}
\end{figure}

\begin{figure}
\begin{center}
\includegraphics[width=85mm]{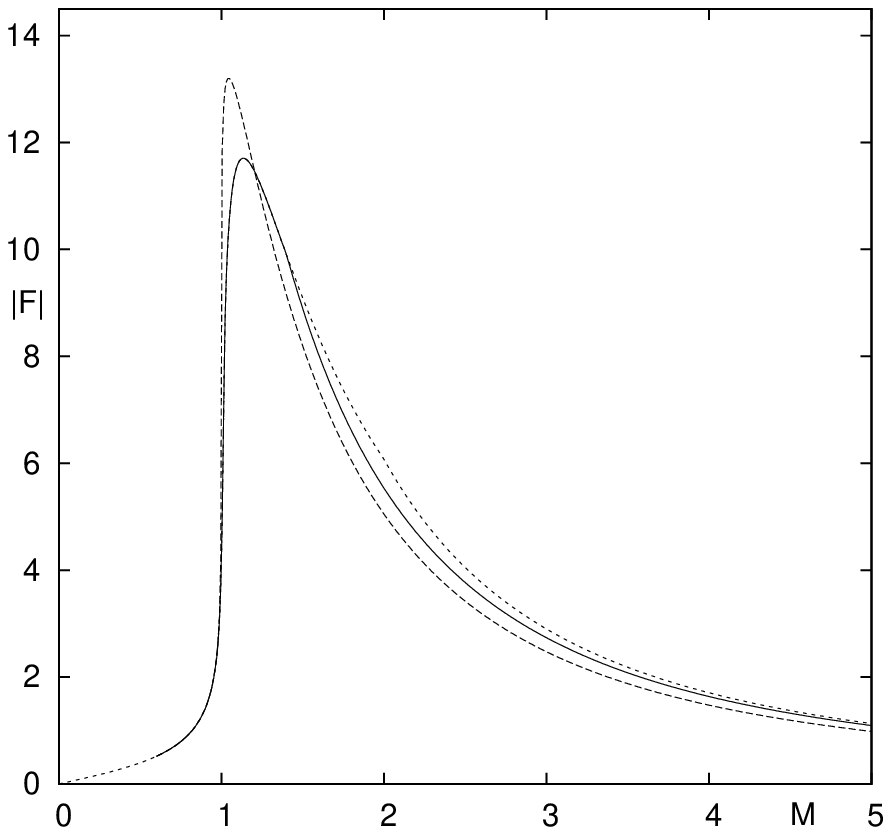}\includegraphics[width=85mm]{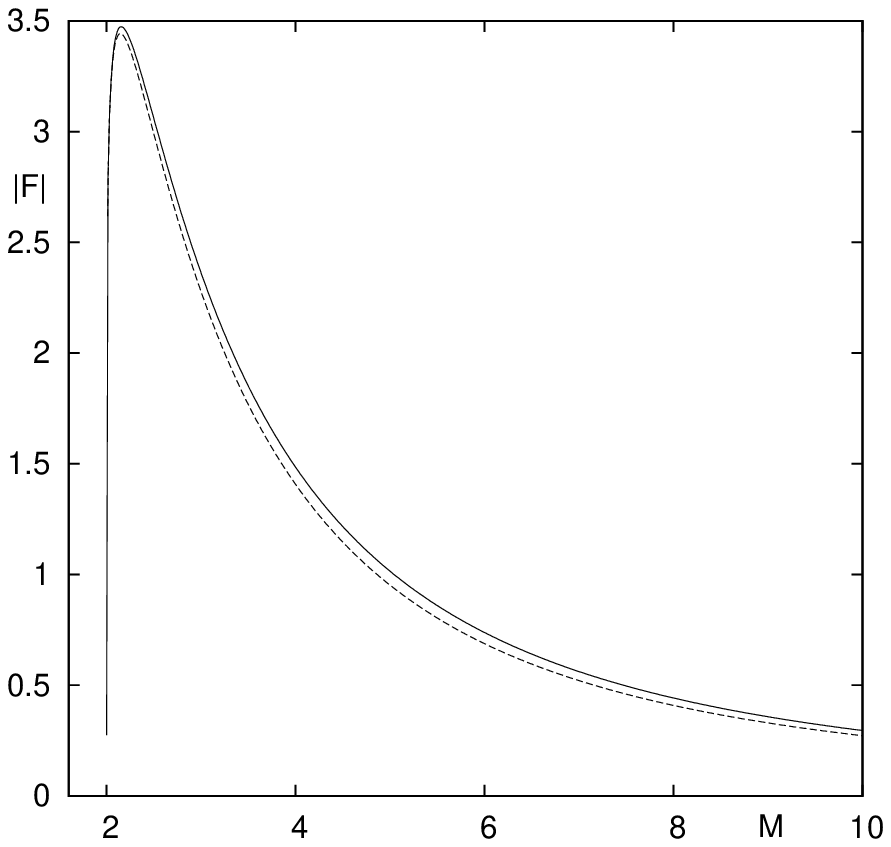}
\end{center}
\caption{Friction force on accelerating perturbers with an initial subsonic velocity (left panel,  ${\cal M}=0.6$) and an initial supersonic velocity (right panel, ${\cal M}_0=2$). The solid curves correspond to the friction force that takes acceleration into account given by equations (\ref{case6}),(\ref{case61}), (\ref{case62}) and (\ref{case63}),  and the dashed curves to the uniform motion friction force (\ref{ostriker}). The small cutoff radius is $r_0=10^{-4}$. The dotted curve in the left panel is the friction force on a perturber initially at rest.}
\label{f4}
\end{figure}

\begin{figure}
\begin{center}
\includegraphics[width=85mm]{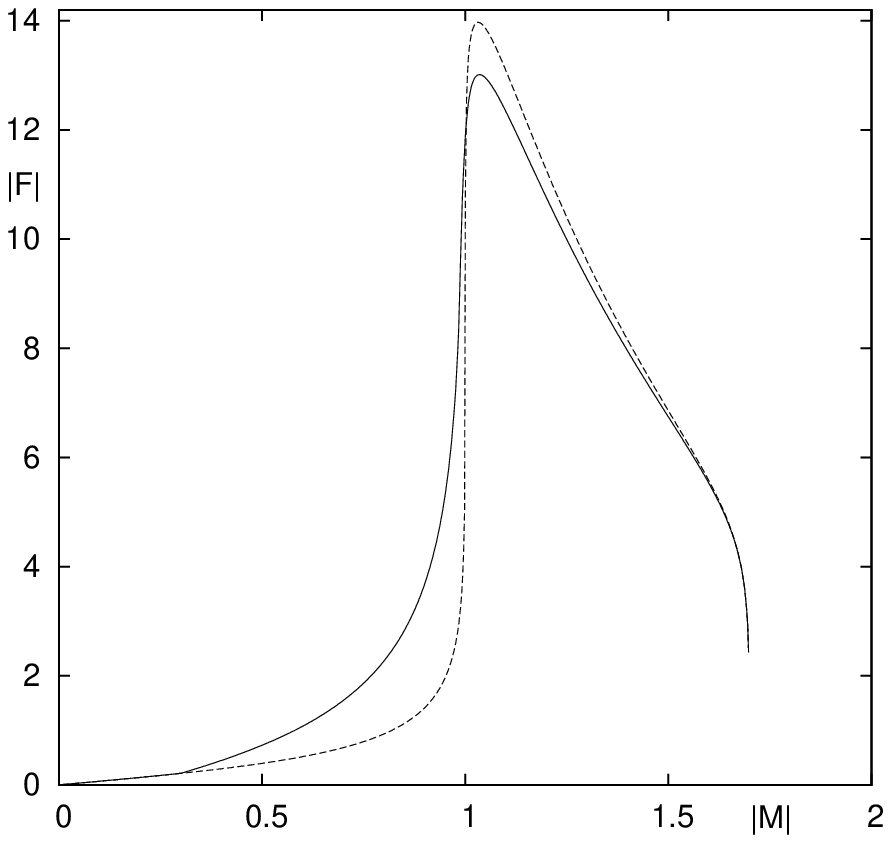}\includegraphics[width=85mm]{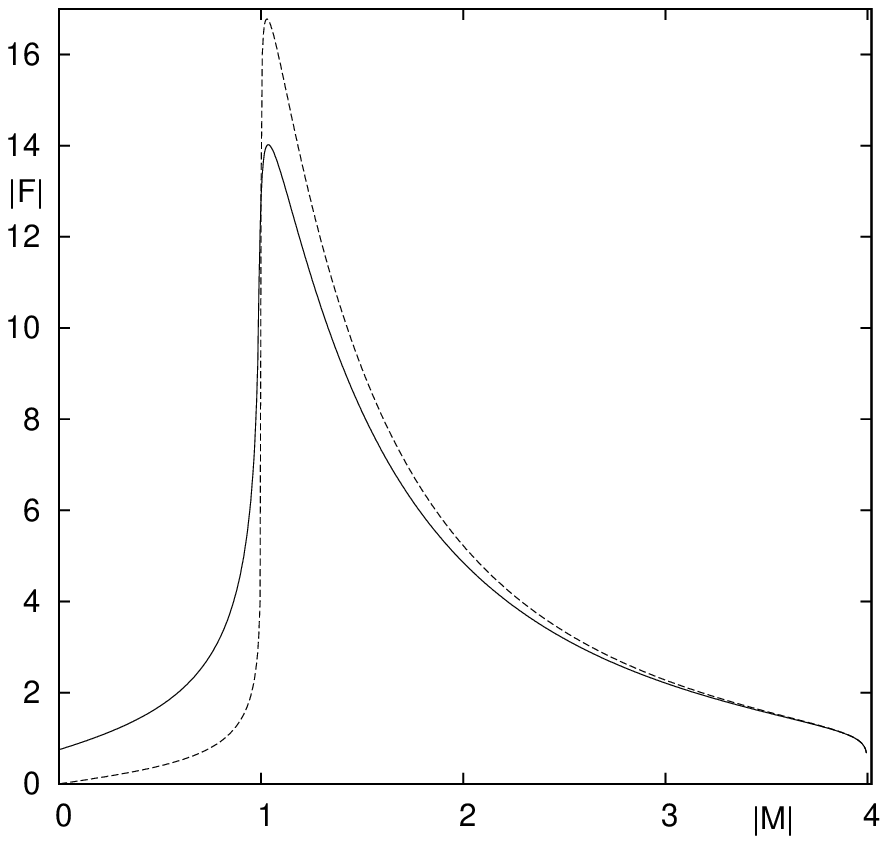}
\end{center}
\caption{Friction force on decelerating perturbers  with initial supersonic velocities  $|{\cal M}_0|=1.7$ (left panel), $|{\cal M}_0|=4$ (right panel). 
The solid curves correspond to the friction force that takes acceleration into account given by equations (\ref{case2}), (\ref{case7}) and (\ref{case8}) for $|{\cal M}_0|=1.7$ and  by equations (\ref{case8x}) and (\ref{case8x2}) for  $|{\cal M}_0|=4$. The dashed curves correspond to the uniform motion friction force (\ref{ostriker}).
The small distance cutoff is $r_0=10^{-4}$.}
\label{f5}
\end{figure}

\begin{figure}
\begin{center}
\includegraphics[width=85mm]{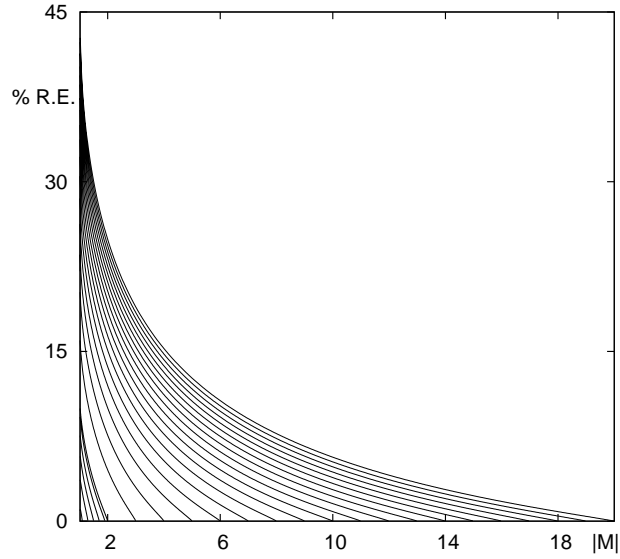}
\end{center}
\caption{Relative error, $100\times |1-|I_0/I_{\rm VII}||$ from equations (\ref{ostriker}) and (\ref{case7}), on the friction force when using the uniform motion expression instead of the accelerated motion expressions for decelerating perturbers with initial supersonic speeds. The small distance cutoff is $r_0=10^{-4}$. The curves correspond to different initial velocities.}
\label{f6}
\end{figure}

\newpage
\begin{figure}
\begin{center}
\includegraphics[width=75mm]{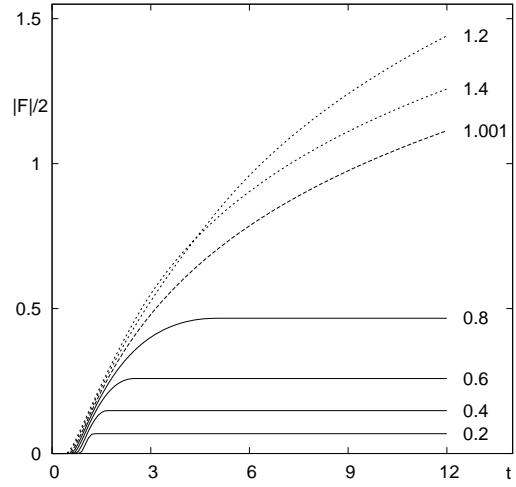}
\end{center}
\label{f7}
\caption{Friction force for uniform motion as a function of time. The force, $F$, is given by (\ref{forc3A}), (\ref{ostriA}), and (\ref{transient}) and  scaled to $2\pi(GM)^2\rho_0/c^2$;  time is scaled to $r_0/c$. The perturber's motion is initiated at $t=0$. The Mach number is indicated to the left of each curve. The time dependence is similar to that obtained from the simulation of the non-linear fluid equations in Figure 2 of S\'anchez-Salcedo and Brandenburg (1999).}
\end{figure}

\end{document}